\documentclass[twocolumn,showpacs,aps,prd,amsmath,amssymb,nofootinbib,nobibnotes,floatfix]{revtex4-1}
\usepackage{hyperref}
\usepackage{amsfonts}
\usepackage{amsmath}
\usepackage{mathrsfs}
\usepackage{epsfig}
\usepackage{graphicx}
\usepackage{url}
\usepackage{hyperref}
\usepackage{float}
\usepackage{color}
\usepackage[utf8]{inputenc} 
\usepackage{graphicx}
\usepackage{epsf}
\usepackage{comment}
\usepackage{slashed}
\usepackage{footmisc}
\usepackage{setspace}
\usepackage[dvipsnames, usenames]{xcolor}
\definecolor{linkcolor}{rgb}{0.0,0.3,0.5}

\newcommand{\tR}{\tilde{R}}

\newcommand{\pa}{{a'}}

\newcommand{\rp}{{{r_{\rm p}}}}

\usepackage{hyperref}
\hypersetup{
    pdfnewwindow=true,      
    colorlinks=true,       
    linkcolor=linkcolor,          
    citecolor=linkcolor,        
    filecolor=blue,      
    urlcolor=blue           
}

\begin{document}

\title{Single-single gravitational-wave captures in globular clusters:\\
Eccentric deci-Hertz sources observable by DECIGO and Tian-Qin}

\author{Johan Samsing$^{1}$, Daniel J. D'Orazio$^{2}$, Kyle Kremer$^{3,4}$, Carl L. Rodriguez$^{2}$, Abbas Askar$^{5}$ }
\affiliation{\vspace{2mm}
$^1$Niels Bohr International Academy, The Niels Bohr Institute, Blegdamsvej 17, 2100 Copenhagen, Denmark.\\
$^2$Department of Astronomy, Harvard University, 60 Garden Street Cambridge, MA 01238, USA. \\
$^3$Department of Physics and Astronomy, Northwestern University, 2145 Sheridan Road, Evanston, IL 60208, USA. \\
$^4$Center for Interdisciplinary Exploration and Research in Astrophysics (CIERA), 2145 Sheridan Road,
Evanston, IL 60208, USA. \\
$^5$Lund Observatory, Department of Astronomy, and Theoretical Physics, Lund University, Box 43, SE-221 00 Lund, Sweden.
}

\begin{abstract}
We study the formation rate of binary black hole mergers formed through gravitational-wave emission between unbound, single black holes in globular clusters. While the formation of these binaries in very dense systems such as galactic nuclei has been well studied, we show here that this process can operate in lower-density stellar systems as well, forming binaries at a rate similar to other proposed pathways for creating eccentric mergers. Recent advances in post-Newtonian cluster dynamics indicate that a large fraction of dynamically-assembled binary black holes merge inside their host clusters during weak and strong binary-single and binary-binary interactions, and that these systems may retain measurable eccentricities as they travel through the LIGO and LISA sensitivity bands.  Using an analytic approach to modeling binary black holes from globular clusters, we show that the formation of merging binaries from previously unbound black holes can operate at a similar rate to mergers forming during strong binary encounters, and that these binaries inhabit a unique region of the gravitational-wave frequency space which can be identified by proposed
deci-Hertz space-based detectors.
\end{abstract}

\maketitle

\section{Introduction}\label{sec:Introduction}

Several binary black hole (BBH) mergers have now been observed  by
LIGO (the Laser Interferometer Gravitational Wave Observatory) and VIRGO, through their emission of gravitational waves (GWs)
\citep{2016PhRvL.116f1102A, 2016PhRvL.116x1103A, 2016PhRvX...6d1015A,
2017PhRvL.118v1101A, 2017PhRvL.119n1101A, 2019arXiv190210331Z, 2019arXiv190407214V}.
However, their astrophysical origin is still unknown, and the observed variety in both BH masses and
spins \cite[e.g.][]{2018arXiv181112907T}, in addition to the observed merger of binary neutron stars (NSs) \citep{2017PhRvL.119p1101A},
indicate that several formation mechanisms might be operating. Some of the recently proposed include:
field binaries \citep{2012ApJ...759...52D, 2013ApJ...779...72D, 2015ApJ...806..263D, 2016ApJ...819..108B,
2016Natur.534..512B, 2017ApJ...836...39S, 2017ApJ...845..173M, 2018ApJ...863....7R, 2018ApJ...862L...3S, 2018MNRAS.480.2011G, 2019arXiv190708297H, 2019MNRAS.485..889S, 2019MNRAS.487....2M, 2019MNRAS.482..870E},
stellar clusters \citep{2000ApJ...528L..17P,
2010MNRAS.402..371B, 2013MNRAS.435.1358T, 2014MNRAS.440.2714B,
2015PhRvL.115e1101R, 2016PhRvD..93h4029R, 2016ApJ...824L...8R,
2016ApJ...824L...8R, 2017MNRAS.464L..36A, 2017MNRAS.469.4665P, 2018MNRAS.480.5645H, 2019MNRAS.487.2947D, 2019ApJ...873..100C, 2019MNRAS.486.3942K, 2019arXiv190611855A},
active galactic nuclei (AGN) discs \citep{2017ApJ...835..165B,  2017MNRAS.464..946S, 2017arXiv170207818M},
galactic nuclei (GN) \citep{2009MNRAS.395.2127O, 2015MNRAS.448..754H,
2016ApJ...828...77V, 2016ApJ...831..187A, 2016MNRAS.460.3494S, 2017arXiv170609896H, 2018ApJ...865....2H},
very massive stellar mergers \citep{Loeb:2016, Woosley:2016, Janiuk+2017, DOrazioLoeb:2017},
and single-single GW captures of primordial black holes \citep{2016PhRvL.116t1301B, 2016PhRvD..94h4013C,
2016PhRvL.117f1101S, 2016PhRvD..94h3504C}.

From the GW signal of individual BBH merger events, one can measure the (redshifted) mass of the BHs, their
spins \citep[e.g.][]{2016ApJ...832L...2R, 2018PhRvD..98h3007N},
the BBH orbital eccentricity \citep[e.g.][]{2017PhRvD..95b4038H, 2018ApJ...855...34G, 2018PhRvD..97b4031H}, and even
Doppler effects related to a possible movement of the BBH's center of mass (COM) \citep[e.g.][]{2017ApJ...834..200M, 2018arXiv180505335R}.
The question is; how can this information be used to distinguish the proposed astrophysical merger channels?
Regarding BH spins, these are expected to be
isotropically distributed for dynamically assembled BBH mergers, such as those forming in globular clusters (GCs), in contrast to those
forming in isolation in the field \citep[e.g.][]{2016ApJ...832L...2R, 2017arXiv170601385F}.
In terms of eccentricity, a significant fraction of BBHs formed dynamically in clusters are expected to lead to a
unique population of GW mergers that have measurable orbital eccentricities in bands from LISA (the Laser Interferometer Space Antenna) \citep{2018MNRAS.tmp.2223S, 2018MNRAS.481.4775D, 2019PhRvD..99f3003K, 2019PhRvD..99f3006S, 2019arXiv190607189S} to LIGO \citep{2006ApJ...640..156G, 2014ApJ...784...71S, 2017ApJ...840L..14S, 2017ApJ...840L..14S, 2018MNRAS.476.1548S, 2018ApJ...853..140S, 2019MNRAS.482...30S, 2018PhRvD..97j3014S, 2018ApJ...855..124S, 2019ApJ...871...91Z, 2018PhRvD..98l3005R}, whereas field binaries are most likely to have fully circularized once observable.
Although eccentric sources might form in other ways, e.g., through Lidov-Kozai oscillations \citep[e.g.][]{2011ApJ...741...82T,2018ApJ...864..134R, 2018arXiv181110627F, 2019arXiv190208604R, 2019arXiv190309160F, 2019MNRAS.486.4443F, 2019arXiv190309659F}, quadruple systems \citep[e.g.][]{2019MNRAS.486.4781F}, and single-single GW
captures \citep[e.g.][]{2009MNRAS.395.2127O, Kocsis:2012ja, 2016PhRvD..94h4013C, 2018ApJ...860....5G}, eccentric BBH mergers
forming in clusters are not only a natural outcome when BHs are present, but can also accurately be modeled using both simple
analytical \citep[e.g.][]{2018PhRvD..97j3014S, 2019arXiv190611855A} and numerical techniques \citep[e.g.][]{2018PhRvD..98l3005R}.
This makes BBH orbital eccentricity a very promising parameter to use for constraining the cluster channel and its astrophysical properties.
In addition to this, it has also been suggested that BBH populations in dense clusters can be probed through their interaction
with nearby stars, e.g., through tidal interactions, which could result in electromagnetic
observables \citep{2019ApJ...877...56L, 2019arXiv190102889S, 2019arXiv190406353K}.

In this paper we continue our studies on how BBH mergers form in stellar clusters, and in particular GCs; systems which
recently have gained significant attention.
One reason is that these systems are relatively easy to model as they stay more or less isolated for almost their entire life,
during which they evolve through clean physical processes involving Newtonian and Post-Newtonian \citep[e.g.][]{2014LRR....17....2B}
$N$-body dynamics. Recent developments in modeling this formation channel include
how BBHs can form through strong binary-single interactions \citep[e.g.][]{2006ApJ...640..156G, 2014ApJ...784...71S, 2017ApJ...846...36S, 2017ApJ...840L..14S, 2018MNRAS.476.1548S, 2018ApJ...853..140S, 2019MNRAS.482...30S, 2018PhRvD..97j3014S, 2018ApJ...855..124S, 2018MNRAS.481.5436S}, weak
binary-single interactions \citep[e.g.][]{2019arXiv190409624H, 2019arXiv190608666H}, strong binary-binary interactions \citep[e.g.][]{2019ApJ...871...91Z}, and
secular interactions \citep[e.g.][]{2019arXiv190201344H, 2019arXiv190201345H, 2019arXiv190700994H}. This recent work especially indicates that
$\sim 50\%$ of all BBHs assembled in GCs are likely to merge inside their cluster with a variety
of GW peak frequencies and eccentricities \citep[e.g.][]{2018PhRvD..98l3005R}, which are tightly connected to the properties of their host cluster, whereas
earlier prescriptions only resolved the dynamically ejected population \citep[e.g.][]{2000ApJ...528L..17P, 2016PhRvD..93h4029R}.
These recent advances have major observational implications when considering the future of GW astrophysics
where planned observatories such as the Einstein Telescope (ET) \citep[e.g.][]{2011CQGra..28i4013H}, and the Cosmic Explorer (CE) \citep[e.g.][]{2017CQGra..34d4001A},
will be able to map out every possible BBH merger in the entire visible Universe.

As one of the last pieces on how BBH mergers might form in GCs through few-body interactions (without the inclusion of a central massive BH),
we here describe the formation of single-single GW capture mergers \citep[e.g.][]{Hansen:1972il, Lee:1993dt, 2018ApJ...860....5G}.
Single-single GW captures have in the literature mainly been associated with GN hosting massive central BHs \citep[e.g.][]{2019ApJ...881...20R}; however,
recent work does indicate that single-single GW captures could operate at a
non-negligible rate in GCs partly due to their relative low velocity dispersion \citep[e.g.][]{2017ApJ...842L...2C}. In our present paper we build upon these earlier studies,
and by the use of analytical arguments we not only prove that single-single GW captures in GCs lead to significant rates, but we also derive how single-single GW captures
form compared to BBH mergers forming through the dominating three-body interaction channel inside the cluster. Especially, we find that
the rate of single-single GW captures relative to the rate of GW mergers forming during binary-single interactions \citep[e.g.][]{2014ApJ...784...71S}, a
population we loosely will refer to as binary-single GW mergers, does not strongly depend on neither the central velocity dispersion nor the density of the cluster,
but mainly on the binary fraction and the shape of the BH population density profile. In most GCs the binary fraction is at the percent level \citep[e.g.][]{2018PhRvD..98l3005R},
for which we find that the rate of single-single GW captures leading to BBH mergers, should be similar to the rate of binary-single GW mergers.

The single-single GW capture mergers distribute differently across GW frequency and orbital eccentricity space
compared to the other before mentioned few-body merger channels. This has been noticed before \citep{2017ApJ...842L...2C}; however, in our
presented paper we derive for the first time the correct normalization of the single-single GW capture mergers
compared to the mergers from binary-single interactions. This allow us to put forward a picture that unifies how BBHs distribute
as a function of the binary fraction and the density profile of the single BH population. We show how the observable GW frequency and orbital
eccentricity distributions change with these properties, which make us propose that single-single GW captures offer unique possibilities to
probe the inner properties of BH subsystems \citep[e.g.][]{2018arXiv181106473A, 2018MNRAS.478.1844A}.

Finally, from a combination of analytical and numerical techniques we illustrate how single-single GW capture mergers form right where the future planned
GW observatories DECIGO \cite{2011CQGra..28i4011K, 2018arXiv180206977I} and Tian Qin \citep{TianQin} are sensitive.
As we clearly demonstrate in this paper, and also previously argued in \citep{2017ApJ...842L...2C},
deci-Hertz observatories fill out a unique gap in GW frequency space that covers the range where the majority of dynamically assembled BBH mergers form, including those from
single-single, binary-single and binary-binary interactions. Such observations will undoubtedly provide unique information about the distribution of
BHs in dense stellar systems. This greatly adds
to the astrophysical motivation to why such detectors should be built.

The paper is organized as follows. We start in Section \ref{sec:GW Inspiral Cross Sections}
by calculating the cross sections for close two-body encounters originating from single-single and binary-single interactions, respectively.
In Section \ref{sec:Rates of GW Inspirals} we convert our cross section expressions to formation rates, and estimate the rate of single-single GW capture mergers
compared to binary-single GW mergers for two different cluster profiles.
We then use simple Monte Carlo (MC) techniques in Section \ref{sec:Observable Implications} to derive GW peak frequency and
eccentricity distributions for our considered dynamical BBH merger channels, and comment on the observational prospects.
Our study is concluded in Section \ref{sec:Conclusions}.

\section{Cross Sections}\label{sec:GW Inspiral Cross Sections}

\begin{figure}
\centering
\includegraphics[width=\columnwidth]{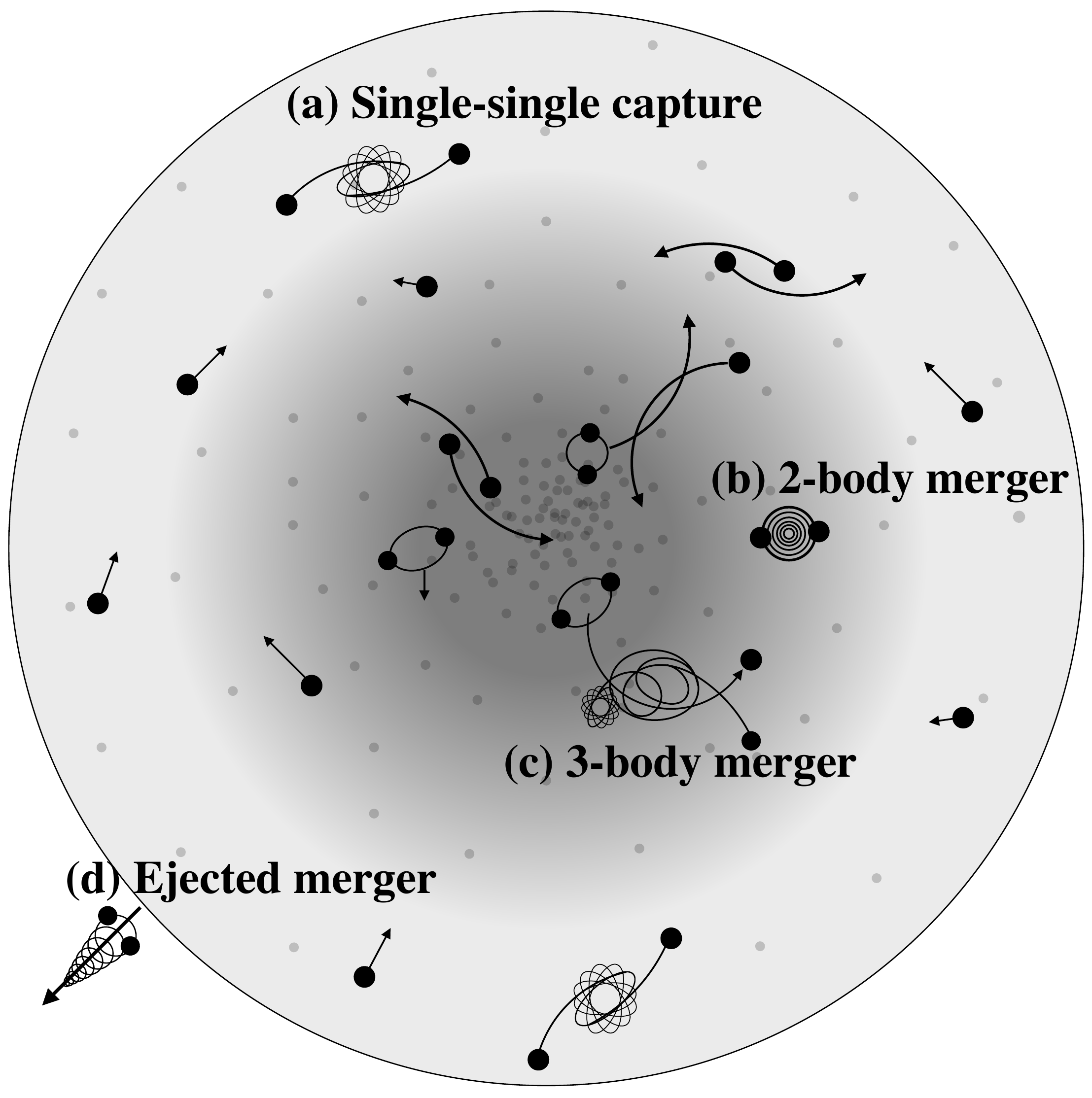}
\caption{Schematic overview of the different BBH merger types that we discuss in this paper.
{\it (a) Single-single capture}: Two initially unbound BHs can become bound if they undergo a passage
small enough for the energy radiated through GW emission to be larger than their initial orbital energy. This
often results in a relatively prompt merger taking place inside the cluster, with a GW peak frequency near the DECIGO band.
BBHs can also form through dynamical processes near the cluster core, and will after formation undergo primarily binary-single
interactions, which result in at least the following three types of merger.
{\it (b) 2-body merger}: If a BBH after a binary-single interaction survives with a SMA and eccentricity such that its GW inspiral time is less than its interaction time, then it will undergo what we refer to as a 2-body merger. The characteristic interaction time scale is $\sim 10^{7}$ years, which maps to a GW peak
frequency near the LISA band.
{\it (c) 3-body merger}: A BBH can also undergo a merger during a binary-single interaction if
its peri-center distance is perturbed to a value small enough for the energy radiated over one orbit
through GW emission is larger than the initial energy of the 3-body system.
The time scale associated with this process is $\sim 1$ year, which maps to a GW peak frequency near the LIGO band.
{\it (d) Ejected merger}: If the BBH does not undergo a 2-body or a 3-body merger inside its cluster,
it will get dynamically ejected through a binary-single interaction. A large fraction of BBHs ejected in this way
will merge outside the cluster within a Hubble time.
}
\label{fig:int_ill}
\end{figure}

In this section we derive close encounter and GW inspiral merger
cross sections for single-single and binary-single interactions, respectively.
These expressions are then used in Section \ref{sec:Rates of GW Inspirals} to derive
absolute and relative merger rates relevant for a system like a GC.
In all our calculations we assume the interacting BHs have the same mass
$m$, which is a reasonable approximation for interactions happening in GCs due to mass segregation
and frequent exchange interactions \citep[e.g.][]{2015ApJ...800....9M, 2018PhRvD..98l3005R}. An illustration showing the interaction- and
GW merger channels we consider in this paper is presented in Fig. \ref{fig:int_ill}. 

In terms of notations, throughout the paper `$G$' denotes Newton's constant, `$c$'
is the speed of light, and `$\log$' denotes the logarithm to the base $10$.

\subsection{Single-Single Interactions}\label{sec:Single-Single Interactions}

We start by considering the cross section for two single BHs to undergo an encounter with peri-center distance ${\rp}$
less than some distance $R_{\rm ss}$. We denote this cross section by $\sigma_{\rm ss}^{<R}$, where `ss' is short for `single-single'.
In the gravitational focusing limit this cross section is given by \citep[e.g.][]{2014ApJ...784...71S, 2018ApJ...853..140S},
\begin{equation}
\sigma_{\rm ss}^{<R} \approx 2 \pi G \frac{2mR_{\rm ss}}{v^2},
\label{eq:sigss_Rs}
\end{equation}
where $m$ is the BH mass, and $v$ is the velocity dispersion of the BH subsystem (we do not distinguish between relative velocity and velocity dispersion in this paper).

For a single-single interaction to also result in a GW capture merger
the energy radiated at the first peri-center passage from GW emission,
$\Delta{E}_{\rm GW} \approx (85\pi/12)G^{7/2}c^{-5}m^{9/2}{\rp}^{-7/2}$ (see \cite{Hansen:1972il}),
must be greater than the initial energy between the two singles, $E_{\rm ss} \approx (1/2) \mu v^{2}$, where $\mu$ is here the
reduced mass \citep[e.g.][]{Lee:1993dt, 2014ApJ...784...71S}.
The maximum peri-center distance from which a single-single GW capture can happen, denoted in this paper by $\mathcal{R}_{\rm ss}$,
is the ${\rp}$ that satisfies $\Delta{E}_{\rm GW} = E_{\rm ss}$, from which one finds,
\begin{equation}
\mathcal{R}_{\rm ss} = \left(\frac{85 \pi}{24\sqrt{2}}\right)^{2/7} \times \mathscr{R}_{\rm m} \left(\frac{c^2}{v^2}\right)^{2/7},
\label{eq:Rcap_ss}
\end{equation}
where $\mathscr{R}_{\rm m}$ is the Schwarzschild radius of a BH with mass $m$. If one substitutes
$R_{\rm ss}$ in Eq. \eqref{eq:sigss_Rs} with $\mathcal{R}_{\rm ss}$ from the above Eq. \eqref{eq:Rcap_ss}
one gets the classical single-single GW capture cross section \citep{Lee:1993dt}, that we here denote by $\sigma_{\rm ss}^{<\mathcal{R}}$.
The value of $\mathcal{R}_{\rm ss}$ given above represents the theoretical upper limit for GW capture; however, the
astrophysical upper limit is slightly smaller, and depends on local properties such as the BH density profile and
velocity dispersion. The primary reason is that the definition of
$\mathcal{R}_{\rm ss}$ given by Eq. \eqref{eq:Rcap_ss} only ensures that the two single objects become bound after their first passage,
but not that this newly formed binary actually undergoes a GW inspiral merger before it is interrupted by a later incoming object.
We will take this correction into account later in our numerical experiments (see Eq. \eqref{eq:Rcaptinsp}), from which we find that this correction for a typical GC is not important.
We therefore keep $\mathcal{R}_{\rm ss}$ as shown in Eq. \eqref{eq:Rcap_ss} throughout our analytical sections.

\subsection{Binary-Single Interactions}\label{sec:Binary-Single Interactions}

We now describe the derivation of the cross section for a binary-single interaction to 
result in two of the three objects to undergo an encounter with peri-center distance ${\rp}$
less than some characteristic distance $R_{\rm bs}$.
We denote this cross section $\sigma_{\rm bs}^{<R}$,
where `bs' is short for `binary-single'. For this, we start by expressing $\sigma_{\rm bs}^{<R}$ as the following product,
\begin{equation}
\sigma_{\rm bs}^{<R} = \sigma_{\rm bs} \times P_{\rm bs}^{<R},
\end{equation}
where $\sigma_{\rm bs}$ is the cross section for a binary to undergo a strong interaction with a single, and $P_{\rm bs}^{<R}$ denotes the
probability for a strong binary-single interaction to result in two of the three objects to undergo an encounter with ${\rp} < R_{\rm bs}$.
The cross section $\sigma_{\rm bs}$ is in the gravitational focusing limit given by \citep{2018ApJ...853..140S, 2018PhRvD..97j3014S},
\begin{equation}
\sigma_{\rm bs} \approx 2 \pi G \frac{3m a}{v^2},
\label{eq:sigma_bs}
\end{equation}
where $a$ is the semi-major axis (SMA) of the target binary.
To derive $P_{\rm bs}^{<R}$ we make use of the formalism described in \citep{2014ApJ...784...71S, 2018PhRvD..97j3014S, 2018MNRAS.tmp.2223S, 2019arXiv190102889S}.
In short, the approach is to split up the chaotic binary-single interaction into a
series of temporary metastable binary-single states, referred to here as intermediate states (IMSs). Each of these $\mathcal{N}$ states can be described
by a binary with a bound single. The eccentricity distribution of the IMS binaries follows approximately that of a so-called
thermal distribution, $P(e) = 2e$ \citep{Heggie:1975uy}, where the SMA is approximately equal to the SMA of the initial target binary, $a$. From this follows that
the probability for a single IMS binary to have a peri-center distance ${\rp}<R_{\rm bs}$ is $ \approx 2R_{\rm bs}/a$ \citep{2018PhRvD..97j3014S}.
As a binary-single interaction on average assembles $\mathcal{N}$ such IMS binaries during its chaotic evolution, the total
probability $P_{\rm bs}^{<R}$ is simply given by,
\begin{equation}
P_{\rm bs}^{<R} \approx \frac{2R_{\rm bs}}{a}\mathcal{N}.
\label{eq:PbsR}
\end{equation}
Using these relations, one now finds that the
cross section $\sigma_{\rm bs}^{<R}$ can be written as,
\begin{equation}
\sigma_{\rm bs}^{<R} \approx  2 \pi G \frac{6mR_{\rm bs}}{v^2} \mathcal{N}.
\label{eq:sigma_bsR}
\end{equation}
From this it is clear that the cross section for a binary-single interaction
to result in a close encounter between two of the three objects
with ${\rp} < R_{\rm bs}$ is $\propto R_{\rm bs}$ and independent of the binary SMA $a$.

As in the single-single case, not all distances $R_{\rm bs}$ in the binary-single problem will lead to a GW inspiral merger.
However, in contrast to the single-single case, defining a unique distance for GW inspiral merger in the binary-single problem is not possible,
as the distance from which a GW inspiral is possible changes for each IMS binary depending on the orbital period of the
remaining bound single \citep{2014ApJ...784...71S}. Therefore, one should in principle work with a distribution of GW inspiral distances; however, this has its own problems, and is therefore out of scope of this paper.
Instead, to simplify our analysis and make it analytically tractable,
we will in this paper work with a single characteristic value for the binary-single GW merger distance, that we
denote by $\mathcal{R}_{\rm bs}$. Note here that this assumption still conserves all the right scaling properties of the problem \citep{2018PhRvD..97j3014S}.
Following \citep{2018PhRvD..97j3014S}, a reasonable value for $\mathcal{R}_{\rm bs}$
is the one for which the GW energy loss integrated over one peri-center passage $\Delta{E}_{\rm GW}$ (see Section \ref{sec:Single-Single Interactions})
for two of the three objects equals the total initial energy of the three-body system $E_{\rm bs}$, that here
is $\approx Gm^{2}/(2a)$. Note here that this requirement is exactly the same as in the single-single case, but with the energy of the initial
binary-single system instead of the energy of the initial single-single system. Now solving for the
peri-center distance ${\rp}$ for which $\Delta{E}_{\rm GW} = E_{\rm bs}$, one finds,
\begin{equation}
\mathcal{R}_{\rm bs} \approx \left(\frac{85 \pi}{24\sqrt{2}}\right)^{2/7} \times {\mathscr{R}_{\rm m}} \left(\frac{a}{{\mathscr{R}_{\rm m}}}\right)^{2/7}.
\label{eq:Rcap_bs}
\end{equation}
Substituting this expression for $\mathcal{R}_{\rm bs}$ into Eq. \eqref{eq:sigma_bsR} results in the cross section for the formation of a GW inspiral that merges
during the binary-single interaction; a cross section we denote $\sigma_{\rm bs}^{<\mathcal{R}}$.
As first noted by \citep{2014ApJ...784...71S}, the cross section $\sigma_{\rm bs}^{<\mathcal{R}} \propto a^{2/7}$, which means that it increases with the initial
SMA $a$. For a thorough discussion on the subject of binary-single interactions with
dissipative terms (GW emission, tidal dissipation, etc.) see \citep[e.g.][]{2017ApJ...846...36S, 2018ApJ...853..140S, 2018MNRAS.481.5436S}.
Finally, as for the single-single GW captures, binary-single interactions and GW inspirals forming during such interactions can in principle be broken up before completion
by incoming objects \citep{2015ApJ...808L..25G}. However, this rarely happens for the systems we consider, and this `infrared' correction will therefore not be
discussed further in this paper.

\subsection{Comparing Binary-Single and Single-Single}\label{sec:Comparing Single-Single and Binary-Single}

One of the key questions we explore in this paper is how important
single-single GW capture mergers are compared to GW inspiral mergers forming during binary-single interactions.
To gain insight into this, we start by considering the following cross section ratio,
\begin{equation}
\frac{\sigma_{\rm bs}^{<R}}{\sigma_{\rm ss}^{<R}} = 3\mathcal{N} (R_{\rm bs}/R_{\rm ss}) \approx 60 (R_{\rm bs}/R_{\rm ss}),
\end{equation}
which follows from the use of Eq. \eqref{eq:sigma_bsR} and Eq. \eqref{eq:sigss_Rs}. Note here that for
the last equality we have used a value of $\mathcal{N} = 20$, which follows from numerical experiments \citep{2018PhRvD..97j3014S}.
From this ratio we conclude that a BBH is about $3\mathcal{N} \approx 60$ times more `effective' in forming two-body encounters
with $r_{\rm p}<R_{\rm ss} = R_{\rm bs}$ compared to a single BH. Note here that this efficiency does not depend on any properties
of the cluster or the BBH orbital parameters.

Two values of $R_{\rm bs}$ and $R_{\rm ss}$ that are particular interesting to compare are the distances from which a
GW merger can form.
By the use of Eq. \eqref{eq:Rcap_bs} and Eq. \eqref{eq:Rcap_ss} we find that,
\begin{equation}
\frac{\mathcal{R}_{\rm bs}}{\mathcal{R}_{\rm ss}} \approx \left( \frac{a}{\mathscr{R}_{\rm m}} \frac{v^{2}}{c^2} \right)^{2/7} = \left(\frac{v}{v_{\rm orb}}\right)^{4/7},
\label{eq:RcapbsRcapss}
\end{equation}
where $v_{\rm orb}$ is the internal orbital velocity of the BBH defined here as the relative velocity between the two BHs assuming a circular orbit. As we only
consider the hard binary (HB) limit in this paper ($v_{\rm orb} \gg v$, see e.g. \citep{Heggie:1975uy}), the above
relations imply that ${\mathcal{R}_{\rm bs}}$ will always be  $< {\mathcal{R}_{\rm ss}}$,
which leads to the following inequality for the corresponding GW capture cross
sections: ${\sigma_{\rm bs}^{<\mathcal{R}}}/{\sigma_{\rm ss}^{<\mathcal{R}}} < 3\mathcal{N}$.

Finally, as will be shown later in Section \ref{sec:Comparing Binary and Single Captures}, a highly relevant ratio
to consider is ${\mathcal{R}_{\rm bs, ej}}/{\mathcal{R}_{\rm ss}}$,
where the subscript `ej' states that $\mathcal{R}_{\rm bs}$ is here evaluated at the SMA $a = a_{\rm ej}$, where $a_{\rm ej}$ is the (maximum) SMA from which the BBH will
get ejected from the cluster through a binary-single interaction \citep{2018PhRvD..97j3014S}.
Assuming that each binary-single interaction decreases the SMA of the BBH by a fixed fraction
$\delta$, such that $a \rightarrow \delta a$, then it follows from classical mechanics that $a_{\rm ej} = (1/6)(\phi - 1)Gm/v_{\rm esc}^2$,
where $v_{\rm esc}$ is the escape velocity of the cluster, and $\phi \equiv 1/\delta$ \citep{2018PhRvD..97j3014S}.
By substituting this relation for $a_{\rm ej}$ into Eq. \eqref{eq:RcapbsRcapss}
one finds,
\begin{equation}
\frac{\mathcal{R}_{\rm bs, ej}}{\mathcal{R}_{\rm ss}} = \left( \frac{\phi - 1}{12 f_{\rm ed}^{2}} \right)^{2/7},
\label{eq:RejRss}
\end{equation}
where $f_{\rm ed} \equiv v_{\rm esc}/v$ (the subscript `ed' refers to that $f$ is the fraction between the ejection (e) and the dispersion (d) velocities, respectively).
Assuming $f_{\rm ed} = 5$, and $\delta = 7/9$ \citep[e.g.][]{Heggie:1975uy, 2018PhRvD..97j3014S}, one finds that ${\mathcal{R}_{\rm bs, ej}}/{\mathcal{R}_{\rm ss}} = 1050^{-2/7} \approx 7.3^{-1}$.
For comparison, if we evaluate this ratio at the HB limit value of the SMA, $a_{\rm HB} = (3/2)Gm/v^2$, which 
is the SMA at which the total energy of the binary, $-Gm^2/(2a)$, equals the energy of incoming single encounters, $(1/2){\mu}v^{2}$, we find
${\mathcal{R}_{\rm bs, HB}}/{\mathcal{R}_{\rm ss}} \approx 1$. Note that this also follows directly from Eq. \eqref{eq:RcapbsRcapss} when $v \approx v_{\rm orb}$.
For our chosen values it therefore follows that
$8 \lesssim {\sigma_{\rm bs}^{<\mathcal{R}}}/{\sigma_{\rm ss}^{<\mathcal{R}}} \lesssim 60$.

\section{Rates}\label{sec:Rates of GW Inspirals}

In this section we convert the cross sections derived in the above
Section \ref{sec:GW Inspiral Cross Sections} to
formation rates. We assume steady state in all our calculations, which of course is a
simplification of how real clusters evolve; however, this assumption
allows us to explore the rates and observables in closed form expressions, which provides crucial guidance to
what to focus on and include in more sophisticated numerical simulations. Also, this allows for
an easier comparison to the recent analytical literature on both binary-single and single-single GW mergers
forming in both GN and GCs.
We proceed below by first deriving a few general relations,
after which we apply these to study the binary-single and single-single rates of GW capture mergers
from two different cluster models; a uniform density model and the Plummer's sphere model.
Corresponding observables, such as the GW peak frequency and BBH orbital
eccentricity will be discussed in Section \ref{sec:Observable Implications}.

\subsection{General Relations}

In the following we present general relations for
deriving absolute and relative rates of GW mergers
resulting from single-single and binary-single interactions, respectively.

\subsubsection{Single-Single Rates}

We consider a spherical cluster with mass density profile $\rho (r)$ and corresponding number density profile $n(r) = \rho (r)/m$,
consisting of identical compact objects with mass $m$.
In a radial shell with width $dr$ at radial position $r$, the number of single-single encounters with ${\rp} < R_{\rm ss}(r)$ per time interval $dt$ is given by,
\begin{equation}
{d\Gamma}_{\rm ss}(r) = n(r) \sigma_{\rm ss}^{<R}(r) v(r) \times n(r) 4 \pi r^{2} dr \times \frac{1}{2},
\label{eq:dGamma_ss}
\end{equation}
where the first term is the rate of encounters with ${\rp} < R_{\rm ss}(r)$ a single object experiences,
the second term is the number of single objects within the considered shell, and the third term corrects for that
the singles are both targets and encounters (two singles result in one encounter). From here
we do not explicitly show if a quantity $x$ is dependent on, e.g., $r$ by writing $x(r)$. This is done to limit the length of our expressions.
Using the above relation from Eq. \eqref{eq:dGamma_ss}, the total single-single rate of encounters with ${\rp} < R_{\rm ss}$ from the entire cluster is now given by,
\begin{equation}
\Gamma_{\rm ss} = \frac{8 \pi^{2} G}{m} \int_{0}^{\infty} \frac{R_{\rm ss} \rho^{2}}{v} r^{2} dr.
\end{equation}
Note here that we have allowed for the characteristic encounter distance $R_{\rm ss}$ to be a function of $r$.
A variable $R_{\rm ss}$ will be important to include when deriving the total rate of single-single GW capture mergers.
The above equation can also be written in the following form,
\begin{equation}
\begin{split}
\Gamma_{\rm ss} &  = n_{0} \sigma_{\rm ss,0}^{<R} v_{0} \frac{1}{2} N_{\rm s} \times \frac{4}{3}\frac{\pi r_{\rm s}^{3} \rho_{0}}{M} \int_{0}^{\infty} \frac{\tilde{R}_{\rm ss} \tilde{\rho}^{2}}{\tilde{v}} 3{\tilde{r}}^{2} d{\tilde{r}}\\
			     &  = n_{0} \sigma_{\rm ss,0}^{<R} v_{0} \frac{1}{2} N_{\rm s} \times \xi_{ss},
\end{split}
\label{sec:Gamma_ss}
\end{equation}
where $N_{\rm s}$ is the total number of singles, $M = N_{\rm s} \times m$ is the total mass,
$\tilde{R}_{\rm ss} = R_{\rm ss}/R_{\rm ss, 0}$, $\tilde{\rho} = \rho/\rho_{0}$, $\tilde{v} = v/v_{0}$, $\tilde{r} = r/r_{\rm s}$, and
\begin{equation}
\xi_{ss} \equiv \frac{4}{3}\frac{\pi r_{\rm s}^{3} \rho_{0}}{M} \int_{0}^{\infty} \frac{\tilde{R}_{\rm ss} \tilde{\rho}^{2}}{\tilde{v}} 3{\tilde{r}}^{2} d{\tilde{r}}.
\label{eq:xi_ss}
\end{equation}
Here the subscript `$0$' denotes the corresponding quantity has to be evaluated at $r=0$ (central cluster values), and $r_{\rm s}$ is a characteristic scale.
It is convenient to consider this notation of the rate, as $\xi_{ss} =1$ for a simple uniform sphere with constant $R_{\rm ss}$. In other words,
the last term, $\xi _{ss}$, represents essentially an `efficiency factor' that depends on how the $N_{\rm s}$ objects distribute within the cluster.

\subsubsection{Binary-Single Rates}\label{sec:Binary-Single Rates}

We now turn to a derivation of the rate of close two-body encounters forming during binary-single interactions.
To analytically calculate this we here make use of the simple dynamical model described in \citep{2018PhRvD..97j3014S, 2019arXiv190102889S}. In short, 
we assume that all binaries form with a SMA equal to their HB value, $a_{\rm HB}$ (see Section \ref{sec:Comparing Single-Single and Binary-Single}),
after which a given binary undergoes interactions with incoming singles that each decreases the binary SMA from $a \rightarrow \delta a$.
This in-cluster hardening process continues until the binary SMA falls below the critical ejection value
$a_{\rm ej}$ (see Section \ref{sec:Comparing Single-Single and Binary-Single}), at which the binding energy released in a single
binary-single interaction leads to ejection of the binary from its cluster.
The average rate of encounters with $r_{\rm p} < R_{\rm bs}$ forming during binary-single interactions can in this model be approximated by
$\Gamma_{\rm bs} \approx N_{\rm bs}/T_{\rm ej}$, where $N_{\rm bs}$ is the total number of encounters with $r_{\rm p} < R_{\rm bs}$ forming
during the hardening binary-single interactions of the binary from $a_{\rm HB}$ to $a_{\rm ej}$, and $T_{\rm ej}$ is the time it takes for the binary to transition from
$a_{\rm HB}$ to $a_{\rm ej}$.

For estimating this rate, we start by deriving $N_{\rm bs}$.
We do this by first considering the differential version of Eq. \eqref{eq:PbsR}, $dN_{\rm bs} = (2R_{\rm bs}/a)\mathcal{N} dk$,
where $k$ here refers to hardening step $k$, i.e. $a(k) = a_{\rm HB}{\delta}^{k}$.
By changing variable from $k$ to SMA $a$
using the relation $da = -a(1-\delta) dk$ now follows,
\begin{equation}
N_{\rm bs} = \frac{2\mathcal{N}}{1-\delta} \int_{a_{\rm ej}}^{a_{\rm HB}} \frac{R_{\rm bs}}{a^{2}} da,
\label{eq:Nbs}
\end{equation}
where we here, and in the rest of the paper, assume that the probability for a BBH to merge inside the cluster is $< 1$,
which is a reasonable assumption for standard GCs.
As a cross check, if we here substitute $R_{\rm bs}$ with the GW inspiral merger distance $\mathcal{R}_{\rm bs}$ given by Eq. \eqref{eq:Rcap_bs},
then one finds that the number of such mergers evaluates to $N_{\rm bs}^{<\mathcal{R}} \approx 7P_{\rm bs, ej}^{<\mathcal{R}}/(5(1-\delta))$,
where $P_{\rm bs, ej}^{<\mathcal{R}}$ is given by Eq. \eqref{eq:PbsR} evaluated at $a = a_{\rm ej}$.
This relation was found in \citep{2018PhRvD..97j3014S}, which serves as an excellent confirmation of our relations so far. Note here that the `enhancement factor' from including
the whole binary-single hardening sequence, and not only the final SMA ($a=a_{\rm ej}$), evaluates to $7/(5(1-\delta)) = 63/10$.

We now turn to deriving the time interval $T_{\rm ej}$, for which we assume the binaries
distribute near the cluster center, such that they only relate to the cluster properties
through the central values, $n_0, v_0$. 
In this approximation we continue by first use that the time between binary-single interactions
is given by the inverse binary-single encounter rate, $(n_0 \sigma_{\rm bs} v_{0})^{-1}$, which can be converted to
the differential form $dt = (n_{0} \sigma_{\rm bs} v_{0})^{-1} dk$, where $t$ here denotes time.
Using $da = -a(1-\delta) dk$, we can now write the total time it takes for a given binary
to transition from $a_{\rm HB}$ to $a_{\rm ej}$
through binary-single scatterings in the following way,
\begin{equation}
\begin{split}
T_{\rm ej}  & = \int_{a_{\rm ej}}^{a_{\rm HB}} \frac{1}{n_0\sigma_{\rm bs}v_0}\frac{da}{a({1-\delta})}, \\
            & \approx \frac{(6 \pi G)^{-1}}{(1-\delta)} \frac{v_0}{n_0} \frac{m^{-1}}{a_{\rm ej}},
\end{split}
\label{eq:Tej}
\end{equation}
where we have used Eq. \eqref{eq:sigma_bs}, and for the last term assumed that $a_{\rm HB} \gg a_{\rm ej}$.
We note here that this is also approximately the inverse rate of binary ejections, i.e., $\Gamma_{\rm ej} \approx 1/T_{\rm ej}$.

By the use of Eq. \eqref{eq:Nbs} and Eq. \eqref{eq:Tej}, and that $\Gamma_{\rm bs} \approx N_{\rm bs}/T_{\rm ej}$,
we now finally find,
\begin{equation}
\begin{split}
\Gamma_{\rm bs} &  = n_{0} \sigma_{\rm bs,ej}^{<R} v_{0} N_{\rm b} \times \int_{1}^{a'_{\rm HB}} \frac{{R_{\rm bs}'}}{{a'}^2} d{a'} \\
			     &  = n_{0} \sigma_{\rm bs,ej}^{<R}v_{0} N_{\rm b} \times \xi_{bs},
\end{split}
\label{eq:Gamma_bs}
\end{equation}
where $N_{\rm b}$ is the number of binaries, $a' = a/a_{\rm ej}$, $R_{\rm bs}' = R_{\rm bs}/R_{\rm bs,ej}$, $\sigma_{\rm bs,ej}^{<R}$ denotes
the cross section evaluated at $a=a_{\rm ej}$, and
\begin{equation}
\xi_{bs} \equiv \int_{1}^{a'_{\rm HB}} \frac{{R_{\rm bs}'}}{{a'}^2} d{a'}.
\label{eq:xi_bs}
\end{equation}
As seen, for a constant value of $R_{\rm bs}'$ the scale-factor $\xi_{bs} \approx 1$ for $a^{\prime}_{\rm HB} \gg 1$. The factor $\xi_{bs}$ therefore
represents the integrated effect from including a scale dependent $R_{\rm bs}$, as e.g. the GW capture distance $\mathcal{R}_{\rm bs}$ given by
Eq. \eqref{eq:Rcap_bs}.
As described in \citep{2018ApJ...853..140S}, the characteristic distance for dissipative captures, including tidal and GW emission captures,
can often be approximated by $R \propto a^{\beta}$, where $\beta$ relates to the energy loss in question, ${\Delta}E$, as ${\Delta}E \propto r_{\rm p}^{-\beta}$. As seen, for this class of `$\beta$-models' the efficiency factor
is given by $\xi_{bs} = (1-\beta)^{-1}$ for $a^{\prime}_{\rm HB} \gg 1$.

\subsubsection{Binary-Single vs. Single-Single}\label{sec:Comparing Binary and Single Captures}

Having derived general forms for the close encounter rate from both single-single, $\Gamma_{\rm ss}$ (Section \ref{sec:Binary-Single Rates}),
and binary-single, $\Gamma_{\rm bs}$ (Section \ref{sec:Binary-Single Rates}), we are now in a position to compare the two. Using Eq. \eqref{eq:Gamma_bs} and
Eq. \eqref{sec:Gamma_ss} we find,
\begin{equation}
\frac{\Gamma_{\rm bs}}{\Gamma_{\rm ss}} \approx 2\frac{N_{\rm b}}{N_{\rm s}} \frac{\sigma_{\rm bs,ej}^{<R}}{\sigma_{\rm ss,0}^{<R}} \frac{\xi_{\rm bs}}{\xi_{\rm ss}}
= {6F_{\rm bs}\mathcal{N}} \times \frac{R_{\rm bs, ej}}{R_{\rm ss,0}} \frac{\xi_{\rm bs}}{\xi_{\rm ss}},
\label{eq:Gammabs_Gammass}
\end{equation}
where the factor $F_{\rm bs}$ denotes the binary
fraction, $F_{\rm bs} = N_{\rm b}/N_{\rm s}$.
This simple expression constitutes one of our main results from this paper. In the following sections we will evaluate this ratio for different characteristic
distances (constant $R$ and GW capture $\mathcal{R}$) and cluster profiles (`MODEL I' and `MODEL II').
The main purpose of this is to illustrate that single-single interactions could contribute with a non-negligible and unique
population of BBH mergers.

\subsection{MODEL I: Uniform Density Sphere}\label{sec:MODEL I: Uniform Density Sphere}

We here consider a BH subsystem described by a uniform density $\rho$ and velocity dispersion $v$.
This example therefore represents a classical `$n \sigma v$' estimate with a finite number of objects, $N_{\rm s}$.
Considering first the single-single population, it is here natural to set $r_{\rm s}$ equal to the size of the cluster, which implies that
$\tilde{\rho} = 1$ and $\tilde{v} = 1$ for $r < r_{\rm s}$, and that $\tR_{\rm ss} = 1$ and $\xi_{ss} = 1$ for a constant $R_{\rm ss}$ and
the GW capture $\mathcal{R}_{\rm ss}$.
For binary-single interactions, the efficiency factor $\xi_{bs}$ from Eq. \eqref{eq:xi_bs} will in contrast only $=1$ for constant $R_{\rm bs}$ encounters,
but not for GW captures with varying $\mathcal{R}_{\rm bs}$.
Below we study the rate of close encounters and GW capture mergers for this model.

\subsubsection{Close Encounters}\label{sec:Close Encounters}

We start by writing out the rate of single-single encounters with $r_{\rm p} < R_{\rm ss}$, where $R_{\rm ss}$ is assumed constant.
Since $\xi_{\rm ss} = 1$, as argued above, then the rate is given by,
\begin{equation}
\Gamma_{\rm ss} \approx n \sigma_{\rm ss}^{<R} v \frac{1}{2} N_{\rm s},
\end{equation}
which follows from Eq. \eqref{sec:Gamma_ss}.
This is not surprisingly the usual rate of encounters per single, `$n \sigma v$', weighted by the total number of singles divided by two, $N_{\rm s}/2$ (see also \citep{2006ApJ...648..411K}).

We now consider the rate of binary-single encounters with $r_{\rm p} < R_{\rm bs}$, where also $R_{\rm bs}$ is here assumed constant.
Since $\xi_{\rm bs} = 1$ also in this case, the rate is simply given by,
\begin{equation}
\Gamma_{\rm bs} \approx n \sigma_{\rm bs}^{<R} v N_{\rm b},
\end{equation}
where we have used Eq. \eqref{eq:Gamma_bs}. Remember here that the cross
section $\sigma_{\rm bs}^{<R}$ is $\propto mR_{\rm bs}/v^2$, and therefore independent of the SMA $a$.

\begin{figure}
\centering
\includegraphics[width=\columnwidth]{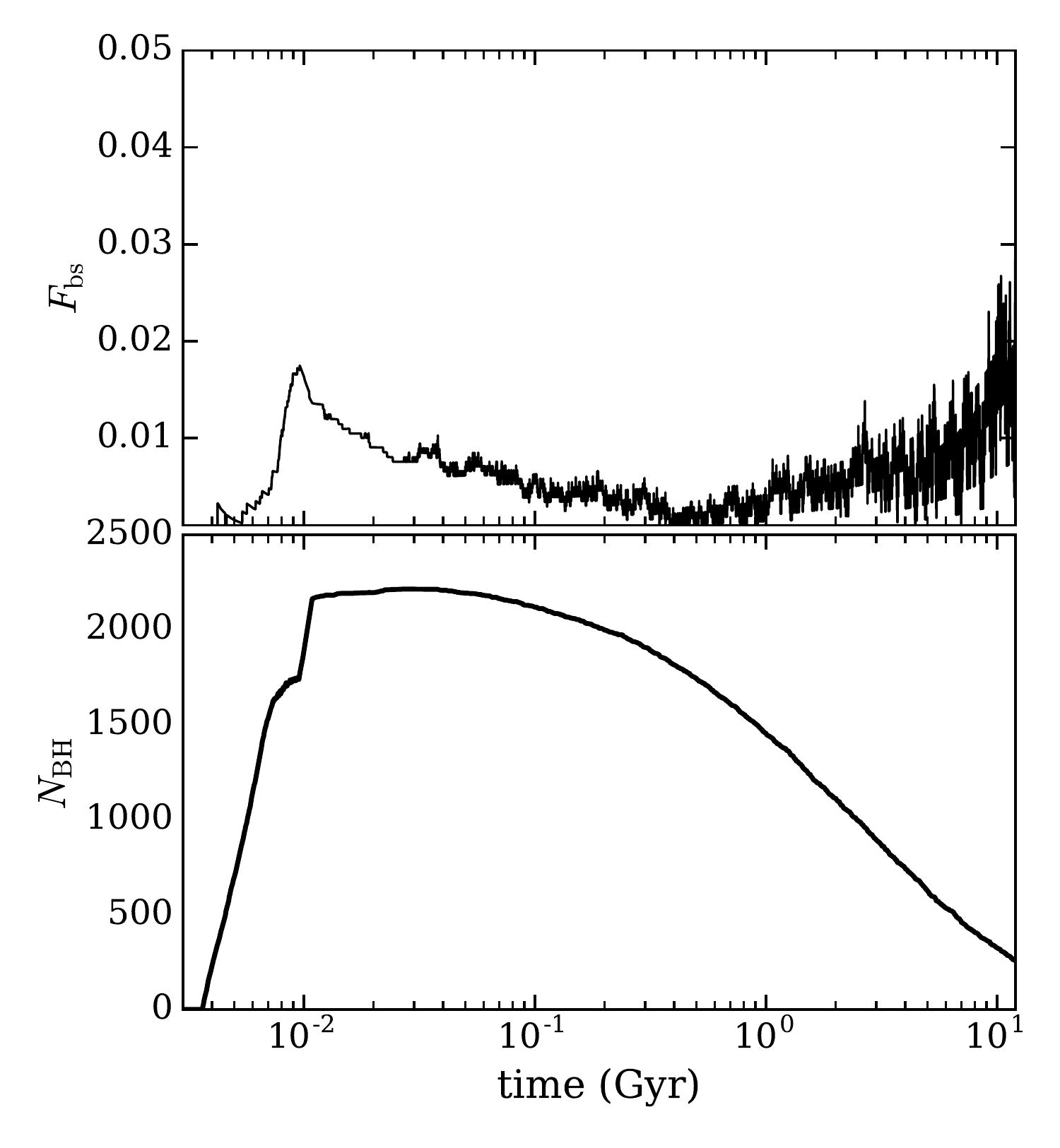}
\caption{Evolution of the BH population in a representative GC (initial conditions: $\sim10^{6}$ stars, $5\%$ binary fraction, and virial radius  $ = 1$ pc),
simulated using the CMC code \citep{2018PhRvD..98l3005R} that is based on a MC Henon solver.
{\it Top}: Number of BBHs relative to the total number of single BHs, $F_{\rm bs} = N_{\rm b}/N_{\rm s}$, as a function
of time after the formation of the GC.
{\it Bottom}: Total number of single BHs in the GC as a function of time. The initial rise is due to single and binary star evolution,
where the later decline arises from dynamical interactions that both eject single BHs from the cluster, and swap them into binaries that eventually
merge. As seen, the binary fraction stays at the percent level throughout the entire life of the GC, which implies that the rate of single-single GW capture mergers
are likely to be similar to the rate of mergers forming during binary-single interactions (see e.g. Eq. \eqref{eq:Gammabs_Gammass_UCR} and Eq. \eqref{eq:xibs_GW}).
However, neither the binary fraction nor the total number of single BHs stays exactly constant, and one therefore expects that the relative rate of single-single GW captures
to change as a function of time. The associated time dependent change of the distribution of GW peak frequency and orbital eccentricity
will make it possible for next generation GW observatories to better constrain the dynamical processes that drive BBHs to merger
in dense stellar systems \citep[e.g.][]{2018ApJ...866L...5R}.}
\label{fig:CMCfig}
\end{figure}

If we now consider the special case for which $R_{\rm ss} = R_{\rm bs}$, then the ratio between
the two rates, $\Gamma_{\rm bs}$ and $\Gamma_{\rm ss}$, reduces to the following expression
\begin{equation}
\frac{\Gamma_{\rm bs}}{\Gamma_{\rm ss}} \approx {6F_{\rm bs}\mathcal{N}},
\label{eq:Gammabs_Gammass_UCR}
\end{equation}
where we have used Eq. \eqref{eq:Gammabs_Gammass}.
This explicitly illustrates that ${\Gamma_{\rm bs}}/{\Gamma_{\rm ss}}$ is independent
of the cluster density, its velocity dispersion, and the orbital evolution of the interacting binaries. Instead, it
essentially only depends on the binary fraction $F_{\rm bs}$. As seen, the two rates are therefore
comparable when the binary fraction $F_{\rm bs} \approx (6\mathcal{N})^{-1}$, from which we conclude that for this model
${\Gamma_{\rm ss}} \approx {\Gamma_{\rm bs}}$ when $F_{\rm bs}  \approx 1/120 \sim 1\%$. 
Therefore, for our considered model, single-single and binary-single encounters with $r_{\rm p} < R_{\rm ss} = R_{\rm bs}$ contribute at the same level
when the binary fraction is at the percent level. Interestingly, this is exactly the level that
is observed in state-of-the-art GC simulations, which explains why single-single interactions might actually contribute with a non-negligible
fraction of BBH mergers. An illustration of this is shown in Fig. \ref{fig:CMCfig}.

As will be discussed later, a constant $R$, as considered here, is important for understanding
the rate of GW sources with a particular GW peak frequency (the GW frequency where most of the power is
outputted) which to leading order only depends on the BBH peri-center distance $r_{\rm p}$ (see Eq. \eqref{eq:fGWdef}). Below we continue by deriving the rate of GW mergers forming within the scale dependent capture distance $\mathcal{R}$.

\subsubsection{GW Mergers}

We now derive the rates of single-single and binary-single mediated GW mergers for our considered uniform cluster model.
One expects a different ratio than the one derived in Eq. \eqref{eq:Gammabs_Gammass_UCR}, as $R$ is now no longer constant
for the binary-single interactions.

We start by deriving the single-single GW capture rate. In this case, the characteristic distance $R_{\rm ss} = \mathcal{R}_{\rm ss}$, but
since the cluster is assumed uniform then $\tR_{\rm ss}$ is still $=1$. The single-single GW capture rate is therefore simply given by 
\begin{equation}
\Gamma_{\rm ss} \approx n {\sigma_{\rm ss}^{<\mathcal{R}}} v \frac{1}{2} N_{\rm s},
\end{equation}
where ${\sigma_{\rm ss}^{<\mathcal{R}}}$ is evaluated using Eq. \eqref{eq:sigss_Rs}, and Eq. \eqref{eq:Rcap_ss}.

We now derive the rate of GW inspiral mergers forming during binary-single interactions. From using Eq. \eqref{eq:Rcap_bs}
it follows that $R'_{\rm bs} = (a/a_{\rm ej})^{2/7} = \pa^{2/7}$, from which we now find using Eq. \eqref{eq:xi_bs} that,
\begin{equation}
\xi_{bs} = \int_{1}^{a^{\prime}_{\rm HB}} \frac{{\pa^{2/7}}}{{\pa}^2} d{\pa} \approx \frac{7}{5}.
\label{eq:xibs_GW}
\end{equation}
For this we assumed the limit $a_{\rm HB} \gg 1$. Note here that this result also follows from setting $\beta = 7/2$
in the general relation $\xi_{bs} = (1-\beta)^{-1}$ derived below Eq. \eqref{eq:xi_bs}. With this value of $\xi_{bs}$,
the rate of GW inspiral mergers forming during binary-single interactions is given by,
\begin{equation}
\Gamma_{\rm bs} = \frac{7}{5} n \sigma_{\rm bs,ej}^{<\mathcal{R}} v N_{\rm b}.
\label{eq:Gamma_bs_uniform}
\end{equation}
The scale dependence of $\mathcal{R}_{\rm ss}$ gives
rise to a small increase of $7/5$, compared to just evaluating the rate at the ejection limit $a = a_{\rm ej}$.

Comparing the rates of GW mergers from the binary-single and single-single channels we find,
\begin{equation}
\frac{\Gamma_{\rm bs}}{\Gamma_{\rm ss}} \approx {6F_{\rm bs}\mathcal{N}} \times \left( \frac{\phi - 1}{12 f_{\rm ed}^{2}} \right)^{2/7} \frac{7}{5},
\end{equation}
where we have used Eq. \eqref{eq:Gammabs_Gammass} and Eq. \eqref{eq:RejRss}. As seen, when accounting for all the GW mergers
that form through single-single and binary-single interactions, and not only those for which $r_{\rm p} < R_{\rm ss} = R_{\rm bs}$ as we did in the above
Section \ref{sec:Close Encounters}, the relative GW merger rate from single-single interactions increases by a factor of
$(5/7)1050^{2/7} \approx 5.2$ for $f_{\rm ed} = 5$ and $\delta  = 7/9$. Therefore, in this case
${\Gamma_{\rm ss}} \approx {\Gamma_{\rm bs}}$ for $F_{\rm bs}  \approx 5.2/120 \sim 5\%$. This indicates that
single-single GW captures very well could play a role in the formation of eccentric in-cluster mergers, 
as $F_{\rm bs}$ is likely $< 5\%$ for standard GCs, as shown by Fig. \ref{fig:CMCfig}.
We proceed below by exploring how our results from this section change when considering a more realistic density profile for the single BH population.

\subsection{MODEL II: Plummer's Sphere}

Real clusters are not described by the simple uniform density sphere that we considered in the above Section \ref{sec:MODEL I: Uniform Density Sphere}.
Instead, relaxation processes generally drive systems to a state described by a profile having a high density in the center and a low in the outskirts \citep[e.g.][]{Spitzer:1969jx}.
To study the effect of a more realistic profile, we here derive absolute and relative rates assuming the single BHs distribute according
to the well-known Plummer's sphere \citep{1911MNRAS..71..460P}. We choose this as it allows for a full analytical treatment of the problem in contrast to other families of profiles.

To start, we first introduce the mass density profile of the Plummer's sphere, which is given by,
\begin{equation}
\rho = \frac{3M}{4 \pi b^{3}} \left( 1 + \frac{r^{2}}{b^{2}}\right)^{-5/2}, 
\end{equation}
and the corresponding velocity dispersion,
\begin{equation}
v^{2} = \frac{1}{6}\frac{GM}{b} \left( 1 + \frac{r^{2}}{b^{2}}\right)^{-1/2},
\label{eq:v2_PLUMMER}
\end{equation}
where $M$ is the total cluster mass, and $b$ is a characteristic scale \citep{1911MNRAS..71..460P}. Inserting these two expressions
into Eq. \eqref{eq:xi_ss}, one finds that the single-single efficiency factor for this profile reduces to the following form
\begin{equation}
\xi_{ss} = 3 \int_{0}^{\infty} \tR_{\rm ss} (1+x^{2})^{-19/4} x^{2} dx.
\label{eq:xi_ss_PLUMMER}
\end{equation}
The binary-single efficiency factor $\xi_{bs}$ is unchanged from what was found in
the above Section \ref{sec:MODEL I: Uniform Density Sphere}, as we assume the binary-single
encounter rate only depends on the central properties of the cluster.
Finally, from using that the potential of the Plummer's sphere is
given by $\phi(r) = -GM/(b\sqrt{1+(r/b)^2})$ and that $v_{\rm esc}(r) = \sqrt{-2\phi(r)}$
it directly follows that $v_{\rm esc,0}/v_{0} = \sqrt{12} \approx 3.5$. As this is very close to our
`fiducial' chosen value of $f_{\rm ed}$, we will still be using $f_{\rm ed} = 5$ in the following
sections to make comparisons more clear.
Below we consider the absolute and relative rate of close encounters and GW mergers for the Plummer's sphere.

\subsubsection{Close Encounters}\label{sec:Close Encounters PLUMMER}

We start by considering the rate of single-single encounters with $r_{\rm p} < R_{\rm ss}$, where $R_{\rm ss}$ is assumed
constant. In this case $\tR_{\rm ss} = 1$, from which we find by the use of Eq. \eqref{eq:xi_ss_PLUMMER} that
\begin{equation}
\xi_{ss} = 3 \int_{0}^{\infty} (1+x^{2})^{-19/4} x^{2} dx = \frac{3\sqrt{\pi}}{4}\frac{\mathscr{G}(13/4)}{\mathscr{G}(19/4)},
\label{eq:xi_ss_R_PLUMMER}
\end{equation}
where $\mathscr{G}(z) = \int_{0}^{\infty} x^{z-1} e^{-x} dx$ denotes the well-known Gamma function. With this factor it is now straight forward
to derive the single-single rate using Eq. \eqref{sec:Gamma_ss}. The question is now, does the change from a uniform sphere to a more realistic density profile
leads to an increase or a decreases of the single-single close encounter rate? As seen here, for the Plummer's
sphere $\xi_{ss} = ({3\sqrt{\pi}}/{4}){\mathscr{G}(13/4)}/{\mathscr{G}(19/4)} \approx 0.2$, which means that the rate from distributing $N_{\rm s}$ singles
in a Plummer's sphere is about $5$ times smaller than if one distributes them in a uniform sphere.
The resulting ratio between the binary-single and single-single close encounter rates for
$R_{\rm ss} = R_{\rm bs}$ is given by,
\begin{equation}
\frac{\Gamma_{\rm bs}}{\Gamma_{\rm ss}} \approx {6F_{\rm bs}\mathcal{N}} \times \frac{4}{3\sqrt{\pi}}\frac{\mathscr{G}(19/4)}{\mathscr{G}(13/4)},
\end{equation}
where we have used Eq. \eqref{eq:Gammabs_Gammass}, and the above Eq. \eqref{eq:xi_ss_R_PLUMMER}. From this
we conclude that ${\Gamma_{\rm ss}} \approx {\Gamma_{\rm bs}}$ for $F_{\rm bs}  \approx 1/600 \sim 0.15\%$,
which suggests that BH subsystems with a realistic density profile is unlikely to
have single-single captures significantly contributing to encounters
with $r_{\rm p} < R_{\rm ss} = R_{\rm bs}$. Below we study the absolute and relative rate of GW mergers.

\subsubsection{GW Mergers}\label{sec:GW Captures PLUMMER}

For single-single GW capture mergers $R_{\rm ss} = \mathcal{R}_{\rm ss}$, which implies that $\tR_{\rm ss} = (v/v_0)^{-4/7}$, where we have used
Eq. \eqref{eq:Rcap_ss}. Plugging this expression into Eq. \eqref{eq:xi_ss_PLUMMER}, and by the use of Eq. \eqref{eq:v2_PLUMMER}, we find the following
value for the single-single efficiency factor,
\begin{equation}
\xi_{ss} = 3 \int_{0}^{\infty} (1+x^{2})^{-129/28} x^{2} dx = \frac{3\sqrt{\pi}}{4}\frac{\mathscr{G}(87/28)}{\mathscr{G}(129/28)},
\label{eq:xi_ss_GW_PLUMMER}
\end{equation}
which (also) evaluates to $\approx 0.2$. Therefore, the rate of single-single GW capture mergers is also greatly reduced for the Plummer's sphere,
compared to the uniform model. Finally, we can also here compare to the rate from binary-single GW mergers.
By the use of Eq. \eqref{eq:Gammabs_Gammass}, the relation shown in Eq. \eqref{eq:RejRss}, and our derived values
for $\xi_{ss}$ (Eq. \eqref{eq:xi_ss_GW_PLUMMER}) and $\xi_{bs}$ (Eq. \eqref{eq:xibs_GW}), we find
\begin{equation}
\frac{\Gamma_{\rm bs}}{\Gamma_{\rm ss}} \approx {6F_{\rm bs}\mathcal{N}} \times \left( \frac{\phi - 1}{12 f_{\rm ed,0}^{2}} \right)^{2/7} \frac{7}{5} \frac{4}{3\sqrt{\pi}}\frac{\mathscr{G}(129/28)}{\mathscr{G}(87/28)},
\label{eq:GbsGss_cap_PLUMMER}
\end{equation}
where $f_{\rm ed,0} \equiv v_{\rm esc,0}/v_{0}$.
For $\delta = 7/9$ and $f_{\rm ed,0} = 5$, the term after the `$\times$'-sign evaluates to $\approx 1$. We therefore conclude that
even when accounting for all GW mergers that can form through single-single and binary-single interactions, the two rates
are only expected to be similar for $F_{\rm bs}  \approx 1/120 \sim 1\%$. However, as previously described, and also shown in Fig. \ref{fig:CMCfig},
$F_{\rm bs}$ is in fact likely to be at the percent level (see Fig. \ref{fig:CMCfig}).

\section{Observable Implications}\label{sec:Observable Implications}

\begin{figure}
\centering
\includegraphics[width=\columnwidth]{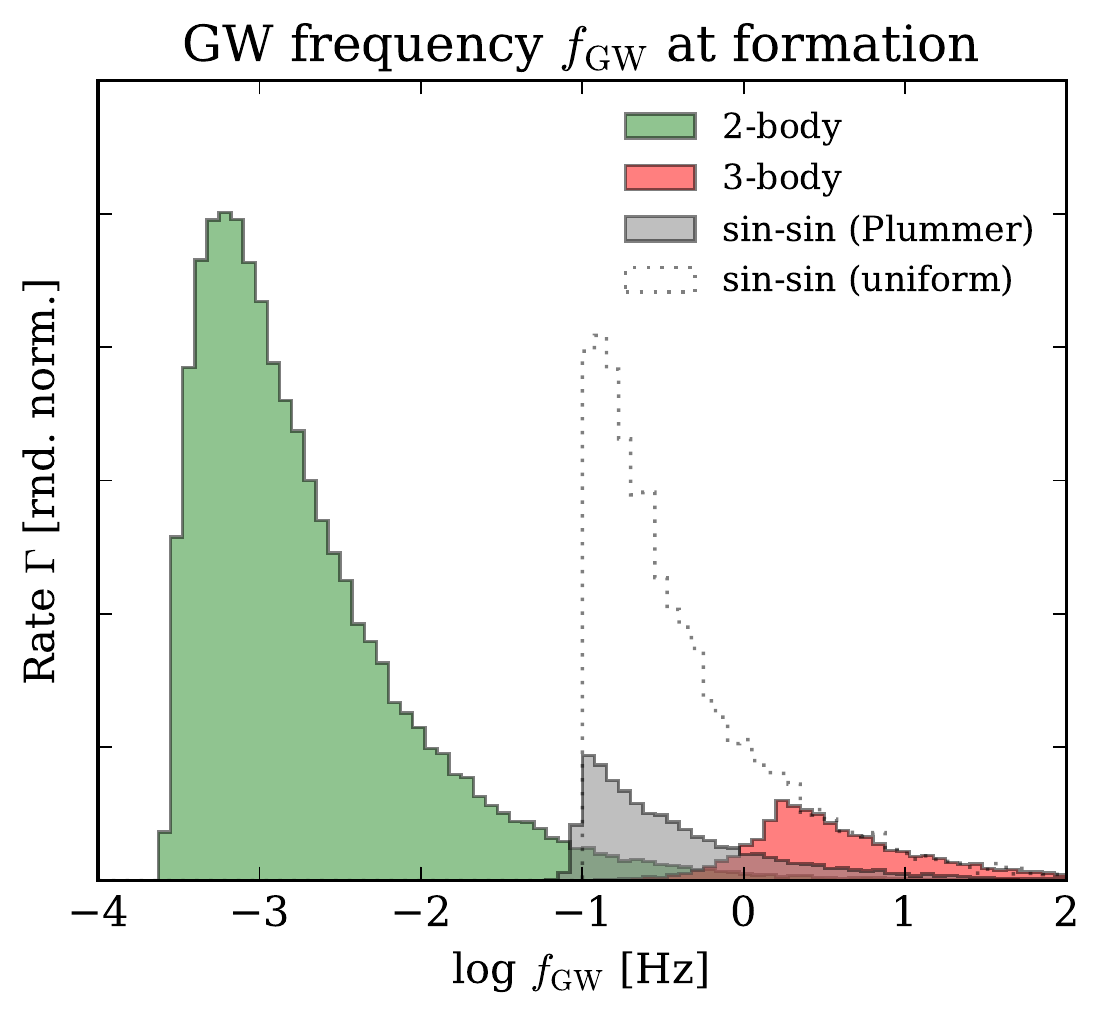}
\caption{Distribution of GW peak frequency ($f_{\rm GW}$) of the BBH mergers forming in our cluster model described in Section \ref{sec:Analytical Monte Carlo Approach}.
Each $f_{\rm GW}$ is derived at the time of formation of the BBH in question, i.e. right when our numerical routine identifies it
as a BBH that will inspiral and merge. Since $f_{\rm GW}$ to leading order only depends on the BBH peri-center distance,
a BBH with a given  $f_{\rm GW}$ will undergo most of its inspiral at that  $f_{\rm GW}$. When a BBH has circularized it will move towards
the right in the above figure, and thereby enter higher frequencies with a lower eccentricity than its initial value.
The `2-body' (green) and `3-body' (red) distributions show results for BBHs merging in-between and during
their binary-single interactions, respectively. The `sin-sin (Plummer)' and the `sin-sin (uniform)' distributions show results
from single-single GW captures forming in a Plummer's sphere and a sphere with a uniform distribution of single BHs, respectively.
As seen, assuming a uniform distribution predicts a surprisingly high fraction of single-single GW capture mergers. Assuming a
more realistic density profile, such as the Plummer's sphere, results in a significant reduction of the rate.
Results are discussed in section \ref{sec:Gravitational-Wave Peak Frequency}, and corresponding eccentricity distributions at
1 Hz and 10 Hz are shown in Fig. \ref{fig:eccdist}.}
\label{fig:fGWdist}
\end{figure}

Having shown using analytical arguments that BH single-single GW captures in GCs might give rise to an observable population of mergers,
we now explore what the observable characteristics are of these mergers. In particular, we explore if single-single GW capture mergers
can be distinguished from the other BBH merger types in GCs. These include BBHs dynamically ejected from their host cluster,
BBHs merging in-between their binary-single interactions, and BBHs merging during their binary-single interactions.
In the following sections we will refer to these GW merger types as {\it ejected mergers}, {\it 2-body mergers}, and {\it 3-body mergers}, respectively,
to shorten labels and descriptions (see also \citep{2019arXiv190607189S}). As pointed
out by \citep[e.g.][]{2017ApJ...842L...2C, 2018MNRAS.tmp.2223S, 2018PhRvD..98l3005R, 2019arXiv190607189S}, these BBH
merger types have different eccentricity- and GW peak frequency distributions. For example, in \citep{2018MNRAS.tmp.2223S}
it was pointed out that 2-body mergers will naturally form near the LISA band with high eccentricity, where 3-body mergers will form at higher frequencies making
them eccentric LIGO sources \citep{2014ApJ...784...71S}.

We here explore how the single-single GW capture mergers will distribute, and in particular what their properties are at 1 Hz and 10 Hz in GW peak frequency. The $1$ Hz regime is potentially interesting for the planned deci-Hertz
observatories DECIGO and Tian Qin, where the 10 Hz limit is naturally interesting for currently operating observatories such as LIGO/VIRGO,
but has also relevance for third generation observatories including ET/CE, which likely will be able to resolve eccentricities
down to $\sim 0.01$ in this range \citep{2015PhRvD..92d4034S}.
It is clear that single-single GW captures are not the dominating source of BBH mergers forming in GCs; however, with future observatories such as ET/CE
we are entering an age where we can expect to see every BBH merger in the visible Universe. Therefore, it is important to have a solid understanding for
how BBH mergers might form in different environments, and what their corresponding observable distributions can tell us about their host systems.
Below we present results from a simple analytical MC method.

\subsection{Analytical Monte Carlo Approach}\label{sec:Analytical Monte Carlo Approach}

To derive BBH merger distributions with the inclusion of single-single interactions, we start by deriving the distribution of 2-body and 3-body mergers
using the semi-analytical MC approach first described in \citep{2018MNRAS.tmp.2223S}. After this, we super impose the distribution from
single-single interactions given some density profile. As for our analytical results, the 2-body and 3-body mergers
are all derived assuming that the binary-single encounter rate is determined by the central quantities of the cluster only. In the following we describe this procedure in detail.

\subsubsection{Modeling Binary-Single Mergers}

For building up the distribution of 2-body and 3-body mergers we follow a large ensemble of uncorrelated BBHs undergoing binary-single interactions
in an environment described by a constant velocity dispersion, $v$, and number density, $n$, equal to the central values of the cluster.
Following \citep{2018MNRAS.tmp.2223S}, each of these BBHs are assumed to form with an initial SMA equal to their hard-binary value
$a_{\rm HB}$; however, for GCs the exact upper value of the SMA
$a$ is not important as long as $a_{\rm HB} \gg a_{\rm ej}$.
Considering now the evolution of one of these BBHs, this BBH will after formation undergo strong binary-single interactions, each of which is assumed to lead
to a constant decrease in the SMA from $a$ to $\delta a$. At the same time, the interactions will also change the eccentricity of the BBH,
where we here assume this change is sampled from the distribution $P(e) = 2e$ \citep{Heggie:1975uy}.
In the point-particle Newtonian limit, this series of binary-single interactions will always end with
a dynamical ejection of the BBH from its host cluster when its SMA falls below $a_{\rm ej}$.
This classical hardening process therefore leads to an ejection of BBHs with a SMA $\sim a_{\rm ej}$,
and an eccentricity distribution that is thermally distributed; however, when GR effects are included in this process, a given BBH can also
merge inside its cluster through (at least) 2-body and 3-body mergers. To account for these two in-cluster
merger types we follow the approach outlined in the paragraphs below. 

{\it 3-body Mergers}: For determining if a BBH merges during a strong 3-body interaction,
we model this often highly chaotic and resonating state using the approach put forward in \citep{2018PhRvD..97j3014S} and briefly
described in Section \ref{sec:Binary-Single Interactions}. In short, the interaction is here divided up into $\mathcal{N}$ IMSs,
each of which is described by a BBH with a bound single BH.
For each IMS we assign the corresponding BBH an eccentricity sampled from the distribution $P(e) =2e$, but keep
the SMA fixed to its initial value $a$. To determine if the BBH undergoes a GW inspiral merger during this IMS, i.e. merge while the
third objects is still bound to it, we compare the energy radiated over one orbit of the BBH ($\Delta{E}_{\rm GW}$),
to the total orbital energy of the bound 3-body state ($E_{\rm bs}$).
If $\Delta{E}_{\rm GW} > E_{\rm bs}$ we label the IMS assembled BBH as a 3-body merger. This energy threshold is equivalent of saying that
the BBH will undergo a GW inspiral merger during the interaction if its peri-center distance $r_{\rm p} < \mathcal{R}_{\rm bs}$.
If instead $r_{\rm p} > \mathcal{R}_{\rm bs}$ the BBH does not merge during the considered IMS.
We repeat this process, i.e. first assigning the IMS assembled BBH an eccentricity from $P(e) = 2e$ and then compare its $r_{\rm p}$ to $\mathcal{R}_{\rm bs}$,
up to $\mathcal{N} =20$ times per interaction.

{\it 2-body Mergers}: If a BBH does not undergo a 3-body merger we determine right after its binary-single interaction if the BBH instead will undergo a 2-body merger. We do this by first assigning an eccentricity to the BBH sampled from $P(e) = 2e$, after which we calculate its inspiral time
$t_{\rm insp}  \propto a^{4}(1-e^2)^{7/2}$ and the time it takes for the next strong single encounter to interact with the BBH, $t_{\rm int} \propto (n_0 \sigma_{\rm bs} v_0)^{-1}$.
If $t_{\rm insp} < t_{\rm int}$ then we label the BBH as a 2-body merger, if instead $t_{\rm insp} > t_{\rm int}$ the BBH survives and we move
on to the next binary-single interaction. 

We repeat this process of checking for 3-body and 2-body mergers while the BBH gradually hardens inside its GC until the BBH either merges
or is ejected. For all 2-body and 3-body mergers we record their orbital parameters, $a$ and $e$, at formation, i.e. 
before they start their inspiral, which allows us to quickly calculate eccentricity and GW peak frequency distributions. These distributions are then
normalized by assuming that the BBHs are formed at a constant rate given by $1/T_{\rm ej}$, where $T_{\rm ej}$ is given by Eq. \eqref{eq:Tej}.

This model clearly represents a simplified picture of how a BH population evolves in a real GC; however, it is very fast, have resulted in very precise
estimates so far, and provides therefore an ideal test-bed for exploring what effects that might be important to included for an accurate modeling
of such systems \citep[e.g.][]{2019arXiv190607189S}.

\begin{figure}
\centering
\includegraphics[width=\columnwidth]{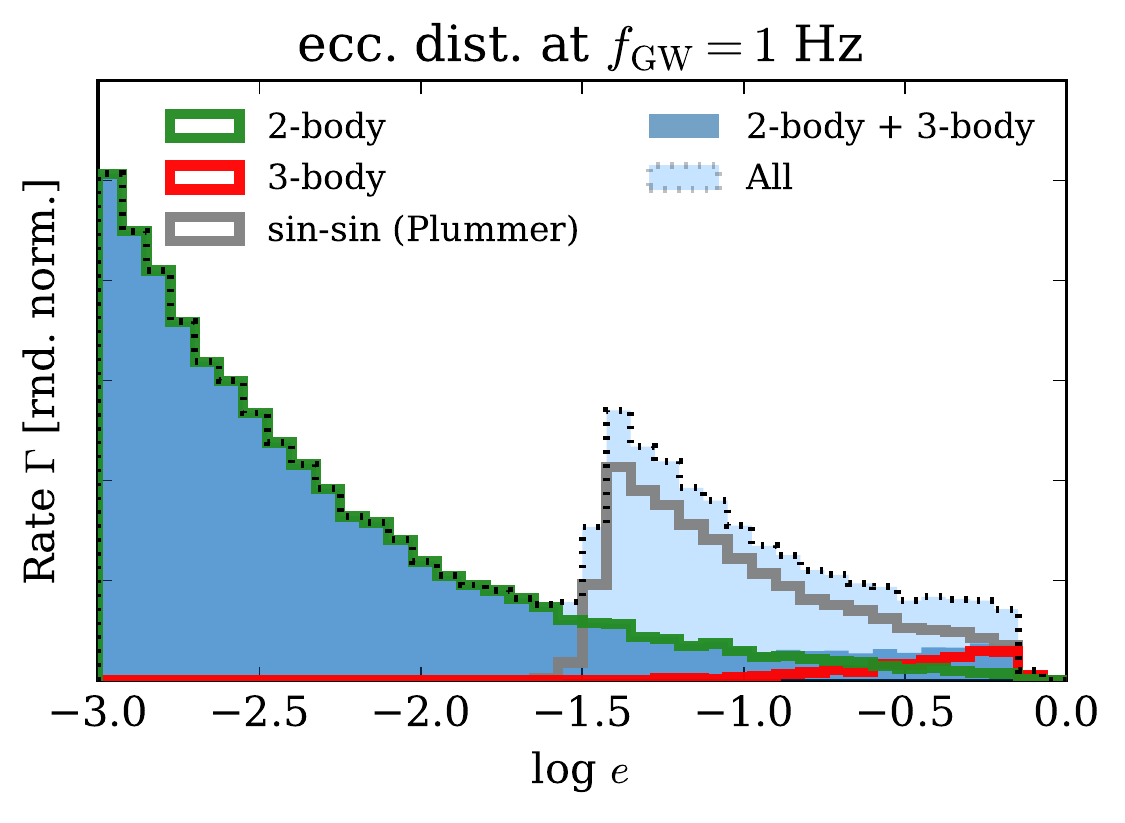}
\includegraphics[width=\columnwidth]{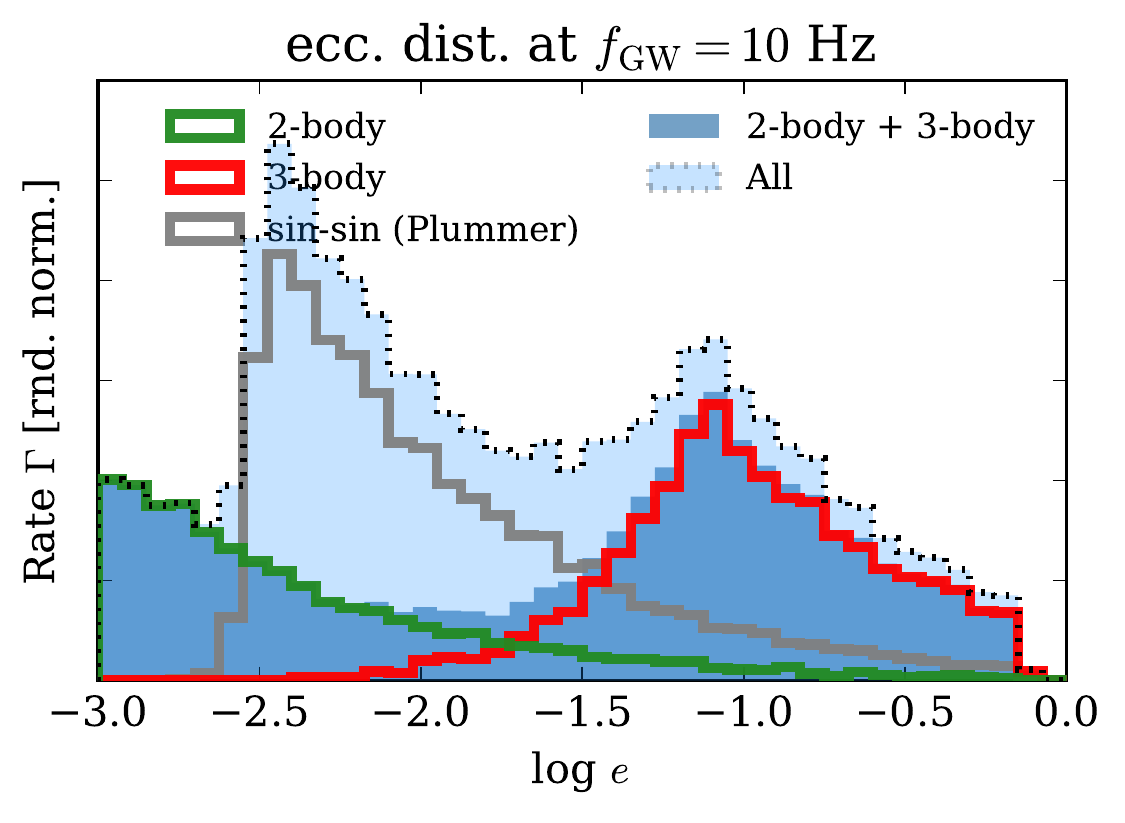}
\caption{Orbital eccentricity distributions of the BBH mergers shown in Fig. \ref{fig:fGWdist} derived from our model
described in Section \ref{sec:Analytical Monte Carlo Approach}. The top and the bottom plots show the distributions at 1 Hz and 10 Hz, respectively. As seen in the top plot,
single-single GW captures dominate the rate of GW sources with eccentricities $\log (e) > -1.5$ among our considered channels at 1 Hz. This regime is relevant for
future planned space borne missions such as DECIGO/Tian Qin.
In the bottom plot is seen that single-single GW captures dominate in the regime $-2.5< \log (e) < 1.5$, which might be resolvable by future ET/CE.
These results depends on the cluster properties (see e.g. \citep{2018PhRvD..97j3014S, 2019arXiv190611855A}), which especially implies
that future detections can be used to constrain the
astrophysical environment of BBH mergers. See Section \ref{sec:Orbital Eccentricity} for further discussions.
}
\label{fig:eccdist}
\end{figure}

\subsubsection{Modeling Single-Single Mergers}

Calculating the single-single GW capture merger distribution is a well defined problem for a non-evolving density profile consisting of equal mass objects,
and can be done in several ways. For this paper we made a small MC routine that simply samples single-single encounters up to the
local GW capture distance $\mathcal{R}_{\rm ss}$.
For each of these encounters we calculate the corresponding orbital parameters, $a = -Gm^{2}(2{\Delta}E_{\rm GW} + 2E_{\rm ss})^{-1}$ and
$e = 1- r_{\rm p}/a$, of the newly formed BBH right after capture, from which we can
determine eccentricity- and GW peak frequency distributions similar to the binary-single case. In this routine, for the local maximum GW capture
distance $\mathcal{R}$ we do not only require that the two single BHs are bound after the encounter, but also that their GW inspiral time,
$t_{\rm insp}(a,e)$, is shorter than the local binary-single interaction time, $t_{\rm int}(a)$.
This requirement results in a local maximum GW capture distance $\mathcal{R}$ that fulfills the following equation,
\begin{equation}
Am^{9/2}\mathcal{R}^{-7/2} - \frac{1}{2}{\mu}v^2 = Bm^{2}\left(\frac{C}{D} \frac{n}{v}\frac{1}{m^{2}}\right)^{2/3} \mathcal{R}^{7/3},
\label{eq:Rcaptinsp}
\end{equation}
where $A = (85{\pi}/12)G^{7/2}c^{-5}$, $B = G/2$, $C = (768/425)(2^{7/2}5/512)G^{-3}c^{5}$,
and $D = (6 \pi G)^{-1}$.
This unfortunately does not have a closed form solution, but can easily be solved numerically. However, we did not
find a significant difference between using the classical capture distance given by Eq. \eqref{eq:Rcap_ss},
and the one found from the above Eq. \eqref{eq:Rcaptinsp}. Results are discussed in the next section.

\subsection{Results}

In this section we present results from numerical experiments, where we follow a population of binary and single
BHs using the routines described in the above Section \ref{sec:Analytical Monte Carlo Approach}. All our results
are based on a system characterized by $m = 20M_{\odot}$, $v_0 = 10 \text{kms}^{-1}$, $n_0 = 10^{5} \text{pc}^{-3}$, $f_{\rm ed,0} = 5$,
and $F_{\rm bs} = 0.01$.

\subsubsection{Gravitational Wave Peak Frequency}\label{sec:Gravitational-Wave Peak Frequency}

We start by considered Fig. \ref{fig:fGWdist}, which shows the GW peak frequency distributions of the merging population of
BBHs right when they form in our MC routine.
For making this, we used the orbital parameters $a,e$ of each of the assembled BBHs to derive their corresponding peri-center distance $r_{\rm p} = a(1-e)$, from which
the GW peak frequency, denoted here by $f_{\rm GW}$, can be approximated by \citep[e.g.][]{Wen:2003bu},
\begin{equation}
f_{\rm GW} \approx \frac{1}{\pi}\sqrt{\frac{2Gm}{r_{\rm p}^3}}.
\label{eq:fGWdef}
\end{equation}
Focusing on the single-single (grey/black) and binary-single (red) mergers we first notice that they peak at different
locations: roughly about $10^{-1}$ Hz and $10^{0.3}$ Hz, respectively.
For both of these channels, their peak location is near the lowest $f_{\rm GW}$
at which all of the mergers from the specific channel is able to contribute (see also \citep{2019PhRvD..99f3006S}). The two peak locations can therefore be expressed
as, $f_{\rm GW,ss}^{\rm peak} \approx {\pi^{-1}}\sqrt{{2Gm}/{\mathcal{R}_{\rm ss,0}^3}} \propto v_0^{6/7}/m$, and
$f_{\rm GW,bs}^{\rm peak} \approx {\pi^{-1}}\sqrt{{2Gm}/{\mathcal{R}_{\rm bs,ej}^3}} \propto v_{\rm esc}^{6/7}/m$, respectively.
Now taking the ratio between these two terms we find the following relation,
\begin{equation}
\frac{f_{\rm GW,bs}^{\rm peak}}{f_{\rm GW,ss}^{\rm peak}} = \left( \frac{12f_{\rm ed,0}^{2}}{\phi - 1}\right)^{3/7}.
\end{equation}
This interestingly shows that the single-single and binary-single peak locations are separated by a constant factor, that in our model
only depends on the properties of the system through $f_{\rm ed,0}$.
For $\delta = 7/9$ and $f_{\rm ed,0} = 5$ follows that ${f_{\rm GW,bs}^{\rm peak}}/{f_{\rm GW,ss}^{\rm peak}} \approx 20$,
which agrees with the results shown in Fig. \ref{fig:fGWdist}. As $f_{\rm ed,0}$ is always $> 1$, our derived ratio further implies that
${f_{\rm GW,bs}^{\rm peak}}/{f_{\rm GW,ss}^{\rm peak}} >  \left({12}/{(\phi - 1)}\right)^{3/7} \approx  5$, therefore, the peaks will always be separated
with ${f_{\rm GW,bs}^{\rm peak}} > {f_{\rm GW,ss}^{\rm peak}}$.

If we now consider the actual shapes of the single-single and binary-single distributions in the region where their GW peak frequencies are greater
than their corresponding distribution peaks, i.e. for $f_{\rm GW} > f_{\rm GW}^{\rm peak}$, we see that they follow the same functional form. This is most easily seen
when comparing the `3-body' with the `sin-sin (uniform)' distributions. The reason is simply that the single-single and the binary-single cross sections
for close encounters with $r_{\rm p} < R$ are both $\propto R$, as seen in Eq. \eqref{eq:sigss_Rs} and Eq. \eqref{eq:sigma_bsR}.
The shape of both of the distributions is therefore given by,
\begin{equation}
\frac{d \Gamma (>f_{\rm GW})}{d\log f_{\rm GW}} \propto f_{\rm GW}\frac{d \sigma (< R)}{df_{\rm GW}} \propto f_{\rm GW}^{-2/3},
\label{eq:dGdlogf}
\end{equation}
where we have used that $\sigma^{<R} \propto R$ for both single-single and binary-single interactions, and the relation $R \propto f_{\rm GW}^{-2/3}$ from Eq. \eqref{eq:fGWdef}. For further discussions on the single-single population and this distribution see \citep{2006ApJ...648..411K}.

Finally, we now consider the relative normalizations of the single-single and binary-single distributions. As seen, for our assumed value of $F_{\rm bs} = 0.01$,
the `sin-sin (uniform)' follows very closely the `3-body' distribution at high values of $f_{\rm GW}$. This is exactly what we derived in Eq. \eqref{eq:Gammabs_Gammass_UCR},
in which we argued that the rates should be similar for encounters with $r_{\rm p} < R$, i.e. for encounters with $f_{\rm GW} > f_{\rm GW}(R)$, when the binary
fraction is at the percent level. This, combined with the $\propto f_{\rm GW}^{-2/3}$ dependence, means that they must
overlap at high $ f_{\rm GW}$. 
The normalizations of the `sin-sin (Plummer)' and `3-body' distributions, over all $f_{\rm GW}$, derived using our analytical methods in Eq. \eqref{eq:GbsGss_cap_PLUMMER}, are also very similar.
To conclude, our analytical and numerical methods agree fully on how and where the different BBH merger populations distribute in $f_{\rm GW}$ space.

\subsubsection{Orbital Eccentricity}\label{sec:Orbital Eccentricity}

We now consider Fig. \ref{fig:eccdist}, which shows the orbital eccentricity distributions of the merging BBHs from Fig. \ref{fig:fGWdist}
at 1 Hz and 10 Hz. The 1 Hz regime is relevant for planned detectors such as DECIGO and Tian Qin, where the 10 Hz
regime is relevant for currently operating LIGO/VIRGO, and future ground-based observatories such as the ET and CE (the ET and CE are likely to be sensitive at even lower GW frequencies).

Starting with the distribution at 1 Hz (top plot), we see that the single-single population (we here only show results for the Plummer's sphere) clearly
dominates the distribution for all eccentricities $\log(e) > -1.5$. That is, in our model, eccentric deci-Hertz sources are likely to originate from single-single
GW captures. This was also noticed in \citep{2017ApJ...842L...2C}; however, we have here been able to derive the correct normalization and how it relates to
the 2-body and 3-body mergers. Note here, that we have in our model not included binary-binary \citep{2019ApJ...871...91Z} and any weak encounter driven mergers
\citep{2019arXiv190607189S}, which in principle also could contribute to eccentric deci-Hertz sources.

If we now consider the distribution at 10 Hz (bottom plot), we see that the single-single GW capture mergers don't
contribute much to the population with an eccentricity resolvable by LIGO/VIRGO ($e>0.1$). Instead, the single-single GW captures seem to fully dominate the
region near $-2.5< \log (e) - 1.5$, which interestingly could possibly be resolved by future detectors such as the ET/CE \citep[e.g.][]{2015PhRvD..92d4034S}.
Therefore, to get a complete picture of what future detectors might observe and how we can use it to constrain the origin of BBH mergers,
the inclusion of single-single GW capture mergers seems to be very important. 

Finally, because highly eccentric orbits will emit qausi-periodic burst of GWs at pericenter, it is observational relevant to consider the timescale between bursts for the single-single and 3-body mergers. As the time between bursts
is equal to the BBH orbital time, $T_{\rm orb}$, the period of interest is simply given by,
\begin{equation}
T_{\rm orb} = \frac{2}{f_{\rm GW}} \left(1-e\right)^{-3/2},
\end{equation}
where we have used the gravitational wave peak frequency, $f_{\rm GW}$ (Eq. \eqref{eq:fGWdef}), together with Kepler's law $T_{\rm orb} = 2 \pi \sqrt{a^{3}/2Gm}$. 
In the high eccentricity limit, i.e. for $e \sim 1$ where bursts are relevant, it is useful to express the eccentricity as $e = 1-10^{-x}$, from which
$T_{\rm orb}$ can be written as $T_{\rm orb} = (2/f_{\rm GW})10^{3x/2}$.
In the low eccentricity limit, i.e. for $e \ll 1$, the orbital period $T_{\rm orb}$ is to leading order given by $T_{\rm orb} \approx (2/f_{\rm GW})(1+3e/2)$.

For representative single-single GW capture mergers that have circularized to the point where they are detectable (near the end of the vertical rise of the grey and red tracks in Figure \ref{fig:fh}), but still on eccentric orbits, e.g., $e \sim 0.9$ and $f_{\rm GW}=0.1$~Hz, $T_{\rm orb, ss} \approx 10$~minutes. For the 3-body capture mergers, in a similar regime, e.g., $e \sim 0.9$ and $f_{\rm GW}=10^{0.3}$~Hz, $T_{\rm orb, bs} \approx 30$~seconds. Work is currently being done on how to detect signals from eccentric burst-like sources \citep[e.g.][]{2017CQGra..34m5011L}.

\subsubsection{Multi-band GW Observations}

\begin{figure}
\centering
\includegraphics[width=\columnwidth]{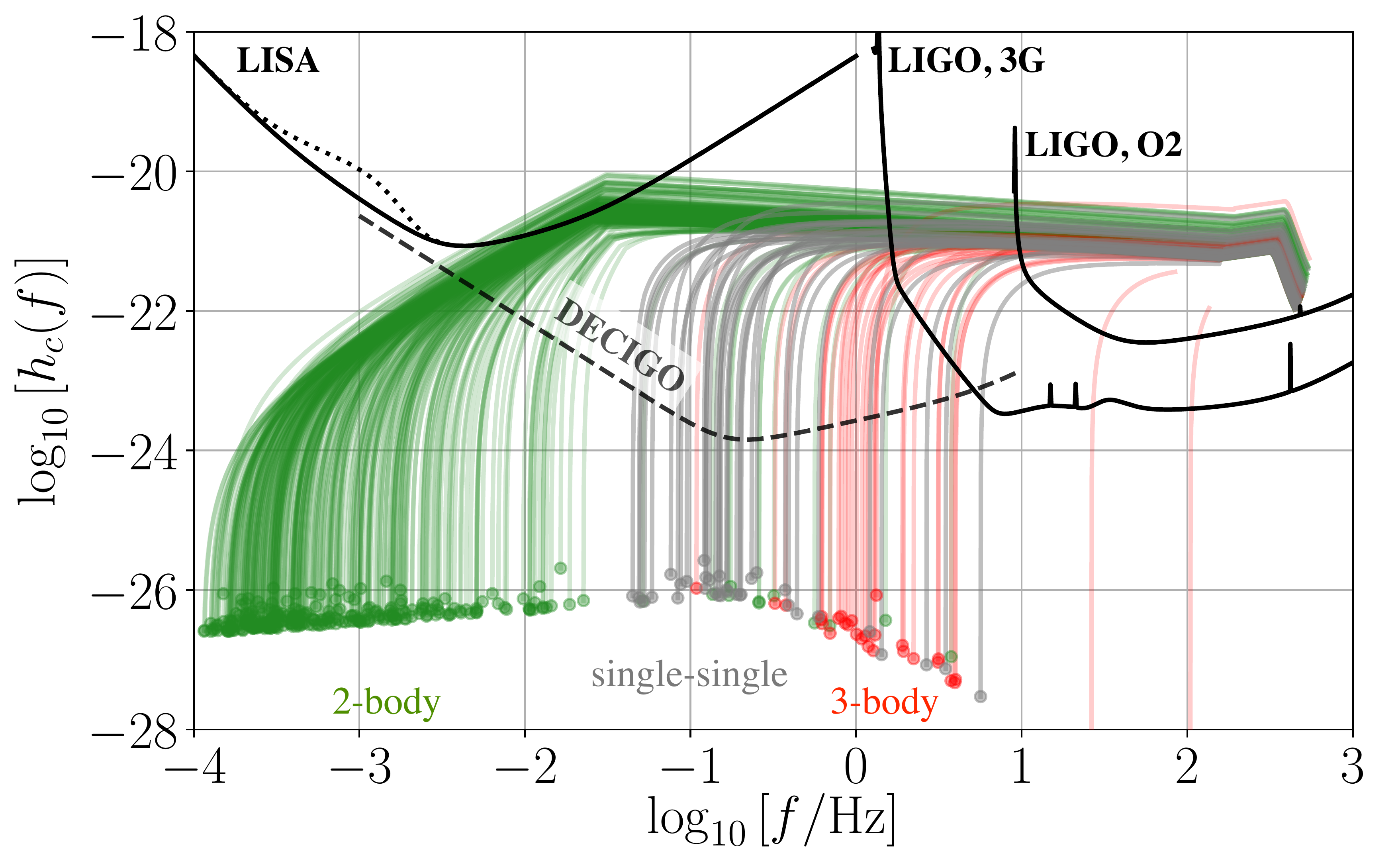}
\caption{Illustration of how our considered 2-body mergers ({\it green}), single-single GW capture mergers ({\it grey}), and
3-body mergers ({\it red}) distribute and evolve as a function of GW peak frequency (x-axis) and GW strain (y-axis). The shown data
is a down-sampled version of the data used for Fig. \ref{fig:fGWdist}.
For the above figure we have assumed the sources distribute uniformly in volume up to a redshift $z = 0.3$.
On the figure is also shown sensitivity curves for LISA, DECIGO, LIGO 3G (the shown curve is for the ET),
and LIGO O2 (current LIGO sensitivity). As seen, the single-single GW capture mergers fill out the gap between the 2-body and 3-body mergers, where DECIGO is most sensitive. The relative contributions from the three plotted merger types change with redshift and cluster parameters, therefore,
future multi-band GW detectors have the potential to reveal exactly how and if BBHs are driven to merger in dense stellar clusters.}
\label{fig:fh}
\end{figure}

We end this section by showing how the 2-body mergers, single-single GW capture mergers, and 3-body mergers distribute and evolve as a function of GW peak frequency
and GW strain to complement recent studies \citep[e.g.][]{2018MNRAS.481.4775D, 2019PhRvD..99f3003K}.
Results are shown in Fig. \ref{fig:fh}, where we have assumed a uniform source population up to a redshift of $z = 0.3$, sampled from the data
shown in Fig. \ref{fig:fGWdist}. On the figure is also shown sensitivity curves for LISA, DECIGO, LIGO 3G (ET), and LIGO O2 (current operational mode).
As is clear from this figure, the single-single GW capture mergers form where DECIGO
is most sensitive, and a future joint multi-band GW network, including instruments like LISA, DECIGO/Tian Qin, and ET/CE, will therefore be able to
put tight constraints on exactly how BHs are brought to merger in dense clusters \citep[e.g.][]{2016PhRvL.116w1102S}.

Note here that the tracks shown in Fig. \ref{fig:fh}
show a representative characteristic strain at any point in the binary lifetime.
They are only corrected for
the observation-time or life-time of the detector in the sense that the characteristic strain is computed differently for binaries that will not significantly evolve in frequency over a chosen four-year observation time. This results in the `knee' in the 2-body tracks (green) that is described more thoroughly in \cite{2018MNRAS.481.4775D}. Binaries for which observation begins anywhere on the track to the left of the knee will not merge within an observation time.
These tracks therefore serve more like an illustrative overview along side with Fig. 1
from \citep{2019PhRvD..99f3003K}. However, we will provide a more in-depth signal-to-noise (S/N) analysis in an upcoming paper (for a discussion on multi-band S/N calculations see \citep[e.g.][]{2016PhRvL.116w1102S, 2018MNRAS.481.4775D}), together with how the observational prospects change with cluster properties and BH masses.

\section{Conclusions}\label{sec:Conclusions}

Recent work on the formation of BBH mergers in GCs has shown that the inclusion of PN few-body dynamics leads to distinctive merger
populations with measurable orbital eccentricities in both
LISA and LIGO \citep[e.g.][]{2018MNRAS.tmp.2223S, 2018MNRAS.481.4775D, 2018PhRvD..98l3005R, 2019PhRvD..99f3003K}.
For these studies, the dynamical few-body channels that so far have been systematically explored include strong- and weak
binary-single interactions \citep[e.g.][]{2014ApJ...784...71S, 2018PhRvD..97j3014S, 2019arXiv190607189S},
and strong binary-binary interactions \citep{2019ApJ...871...91Z}.

In this paper we have expanded on this emerging picture of how BBHs are driven to merger in GCs,
by studying the formation of single-single GW capture mergers. For this we have focused on
deriving the rate of BBH mergers from this single-single channel relative to that of BBH mergers forming during binary-single interactions.
This approach gives us a much better handle on exactly how frequent single-single GW capture mergers might be, in contrast to other studies
that report (highly uncertain) absolute rates.

Using analytical arguments, we find that the contribution from single-single GW
capture mergers relative to those forming during binary-single interactions, does not strongly depend on either the absolute value of the central velocity dispersion,
or the BH number density (see e.g. Eq. \eqref{eq:Gammabs_Gammass}).
Instead, it mainly depends on the number of BBHs relative to single BHs, i.e. the BH binary fraction, and the shape of the density profile of the
single BH population. Assuming the single BHs follow either a uniform- or a Plummer's distribution,
we have shown that the rate of single-single GW capture mergers should be comparable
to that of mergers forming during binary-single interactions when the binary fraction is at the percent level. Interestingly, recent MC simulations
(see Fig. \ref{fig:CMCfig}) do in fact indicate that the fraction of dynamically formed BBHs is at the percent level,
which leads us to conclude that single-single GW capture mergers are expected to form at an observable rate.
Knowing the relative contributions from both single-single and binary-single interactions is extremely important, as their interplay and
resulting GW observables provide the key to probe how BBHs might form in dense stellar clusters.

Finally, using a semi-analytical MC framework we derived GW peak frequency- and eccentricity distributions for
single-single GW capture mergers, and mergers forming during and in-between strong binary-single interactions. As also noticed by \citep{2017ApJ...842L...2C},
the single-single GW capture mergers form right in the deci-Hertz regime where the proposed DECIGO and Tian Qin detectors are sensitive.
For a binary fraction of $F_{\rm bs} = 0.01$ and assuming a Plummer's sphere for the single BH population, we further illustrated that BBHs with eccentricities $\log(e) > -1.5$
at 1 Hz and $-1.5 > \log(e) > -2.5$ at 10 Hz are dominated by the single-single GW capture channel. This has major implications for mapping out how BBHs
form as a function of redshift in dense stellar clusters with DECIGO/Tian Qin as well as third generation GW detectors such as the ET and CE.

{\it Acknowledgments. ---}
It is a pleasure to thank the organizers of the
CIERA Mini-workshop on Post-Newtonian Dynamics in Stellar Clusters, December 5-8, 2018,
where part of this work was initiated. The authors also thank Mirek Giersz for useful
comments on the manuscript.
JS acknowledges support from the Lyman Spitzer Fellowship
and the European Unions Horizon 2020 research and innovation programme under the Marie Sklodowska-Curie
grant agreement No. 844629.
DJD acknowledges financial
support from NASA through Einstein Postdoctoral Fellowship award
number PF6-170151.
KK acknowledges support by the National
Science Foundation Graduate Research Fellowship Program under Grant No. DGE-1324585.
CR acknowledges supported from the Pappalardo Postdoctoral Fellowship at MIT.
AA is supported by the Carl Tryggers Foundation through
the grant CTS 17:113.

\bibliographystyle{h-physrev}
\bibliography{NbodyTides_papers}

\begin{thebibliography}{100}

\bibitem{2016PhRvL.116f1102A}
B.~P. {Abbott} {\em et~al.},
\newblock Physical Review Letters {\bf 116}, 061102 (2016), 1602.03837.

\bibitem{2016PhRvL.116x1103A}
B.~P. {Abbott} {\em et~al.},
\newblock Physical Review Letters {\bf 116}, 241103 (2016), 1606.04855.

\bibitem{2016PhRvX...6d1015A}
B.~P. {Abbott} {\em et~al.},
\newblock Physical Review X {\bf 6}, 041015 (2016), 1606.04856.

\bibitem{2017PhRvL.118v1101A}
B.~P. {Abbott} {\em et~al.},
\newblock Physical Review Letters {\bf 118}, 221101 (2017), 1706.01812.

\bibitem{2017PhRvL.119n1101A}
B.~P. {Abbott} {\em et~al.},
\newblock Physical Review Letters {\bf 119}, 141101 (2017), 1709.09660.

\bibitem{2019arXiv190210331Z}
B.~{Zackay}, T.~{Venumadhav}, L.~{Dai}, J.~{Roulet}, and M.~{Zaldarriaga},
\newblock arXiv e-prints , arXiv:1902.10331 (2019), 1902.10331.

\bibitem{2019arXiv190407214V}
T.~{Venumadhav}, B.~{Zackay}, J.~{Roulet}, L.~{Dai}, and M.~{Zaldarriaga},
\newblock arXiv e-prints , arXiv:1904.07214 (2019), 1904.07214.

\bibitem{2018arXiv181112907T}


\bibitem{2017PhRvL.119p1101A}
B.~P. {Abbott} {\em et~al.},
\newblock Physical Review Letters {\bf 119}, 161101 (2017), 1710.05832.

\bibitem{2012ApJ...759...52D}
M.~{Dominik} {\em et~al.},
\newblock \apj {\bf 759}, 52 (2012), 1202.4901.

\bibitem{2013ApJ...779...72D}
M.~{Dominik} {\em et~al.},
\newblock \apj {\bf 779}, 72 (2013), 1308.1546.

\bibitem{2015ApJ...806..263D}
M.~{Dominik} {\em et~al.},
\newblock \apj {\bf 806}, 263 (2015), 1405.7016.

\bibitem{2016ApJ...819..108B}
K.~{Belczynski} {\em et~al.},
\newblock \apj {\bf 819}, 108 (2016), 1510.04615.

\bibitem{2016Natur.534..512B}
K.~{Belczynski}, D.~E. {Holz}, T.~{Bulik}, and R.~{O'Shaughnessy},
\newblock \nat {\bf 534}, 512 (2016), 1602.04531.

\bibitem{2017ApJ...836...39S}
K.~{Silsbee} and S.~{Tremaine},
\newblock \apj {\bf 836}, 39 (2017), 1608.07642.

\bibitem{2017ApJ...845..173M}
A.~{Murguia-Berthier}, M.~{MacLeod}, E.~{Ramirez-Ruiz}, A.~{Antoni}, and
  P.~{Macias},
\newblock \apj {\bf 845}, 173 (2017), 1705.04698.

\bibitem{2018ApJ...863....7R}
C.~L. {Rodriguez} and F.~{Antonini},
\newblock \apj {\bf 863}, 7 (2018), 1805.08212.

\bibitem{2018ApJ...862L...3S}
S.~L. {Schr{\o}der}, A.~{Batta}, and E.~{Ramirez-Ruiz},
\newblock \apjl {\bf 862}, L3 (2018), 1805.01269.

\bibitem{2018MNRAS.480.2011G}
N.~{Giacobbo} and M.~{Mapelli},
\newblock \mnras {\bf 480}, 2011 (2018), 1806.00001.

\bibitem{2019arXiv190708297H}
A.~S. {Hamers} and T.~A. {Thompson},
\newblock arXiv e-prints , arXiv:1907.08297 (2019), 1907.08297.

\bibitem{2019MNRAS.485..889S}
M.~{Spera} {\em et~al.},
\newblock \mnras {\bf 485}, 889 (2019), 1809.04605.

\bibitem{2019MNRAS.487....2M}
M.~{Mapelli}, N.~{Giacobbo}, F.~{Santoliquido}, and M.~C. {Artale},
\newblock \mnras {\bf 487}, 2 (2019), 1902.01419.

\bibitem{2019MNRAS.482..870E}
J.~J. {Eldridge}, E.~R. {Stanway}, and P.~N. {Tang},
\newblock \mnras {\bf 482}, 870 (2019), 1807.07659.

\bibitem{2000ApJ...528L..17P}
S.~F. Portegies~Zwart and S.~L.~W. McMillan,
\newblock \apj {\bf 528}, L17 (2000).

\bibitem{2010MNRAS.402..371B}
S.~{Banerjee}, H.~{Baumgardt}, and P.~{Kroupa},
\newblock \mnras {\bf 402}, 371 (2010), 0910.3954.

\bibitem{2013MNRAS.435.1358T}
A.~{Tanikawa},
\newblock \mnras {\bf 435}, 1358 (2013), 1307.6268.

\bibitem{2014MNRAS.440.2714B}
Y.-B. {Bae}, C.~{Kim}, and H.~M. {Lee},
\newblock \mnras {\bf 440}, 2714 (2014), 1308.1641.

\bibitem{2015PhRvL.115e1101R}
C.~L. {Rodriguez} {\em et~al.},
\newblock Physical Review Letters {\bf 115}, 051101 (2015), 1505.00792.

\bibitem{2016PhRvD..93h4029R}
C.~L. {Rodriguez}, S.~{Chatterjee}, and F.~A. {Rasio},
\newblock \prd {\bf 93}, 084029 (2016), 1602.02444.

\bibitem{2016ApJ...824L...8R}
C.~L. {Rodriguez}, C.-J. {Haster}, S.~{Chatterjee}, V.~{Kalogera}, and F.~A.
  {Rasio},
\newblock \apjl {\bf 824}, L8 (2016), 1604.04254.

\bibitem{2017MNRAS.464L..36A}
A.~{Askar}, M.~{Szkudlarek}, D.~{Gondek-Rosi{\'n}ska}, M.~{Giersz}, and
  T.~{Bulik},
\newblock \mnras {\bf 464}, L36 (2017), 1608.02520.

\bibitem{2017MNRAS.469.4665P}
D.~{Park}, C.~{Kim}, H.~M. {Lee}, Y.-B. {Bae}, and K.~{Belczynski},
\newblock \mnras {\bf 469}, 4665 (2017), 1703.01568.

\bibitem{2018MNRAS.480.5645H}
J.~{Hong} {\em et~al.},
\newblock \mnras {\bf 480}, 5645 (2018), 1808.04514.

\bibitem{2019MNRAS.487.2947D}
U.~N. {Di Carlo} {\em et~al.},
\newblock \mnras {\bf 487}, 2947 (2019), 1901.00863.

\bibitem{2019ApJ...873..100C}
N.~{Choksi}, M.~{Volonteri}, M.~{Colpi}, O.~Y. {Gnedin}, and H.~{Li},
\newblock \apj {\bf 873}, 100 (2019), 1809.01164.

\bibitem{2019MNRAS.486.3942K}
J.~{Kumamoto}, M.~S. {Fujii}, and A.~{Tanikawa},
\newblock \mnras {\bf 486}, 3942 (2019), 1811.06726.

\bibitem{2019arXiv190611855A}
F.~{Antonini} and M.~{Gieles},
\newblock arXiv e-prints , arXiv:1906.11855 (2019), 1906.11855.

\bibitem{2017ApJ...835..165B}
I.~{Bartos}, B.~{Kocsis}, Z.~{Haiman}, and S.~{M{\'a}rka},
\newblock \apj {\bf 835}, 165 (2017), 1602.03831.

\bibitem{2017MNRAS.464..946S}
N.~C. {Stone}, B.~D. {Metzger}, and Z.~{Haiman},
\newblock \mnras {\bf 464}, 946 (2017), 1602.04226.

\bibitem{2017arXiv170207818M}
B.~{McKernan} {\em et~al.},
\newblock ArXiv e-prints  (2017), 1702.07818.

\bibitem{2009MNRAS.395.2127O}
R.~M. {O'Leary}, B.~{Kocsis}, and A.~{Loeb},
\newblock \mnras {\bf 395}, 2127 (2009), 0807.2638.

\bibitem{2015MNRAS.448..754H}
J.~{Hong} and H.~M. {Lee},
\newblock \mnras {\bf 448}, 754 (2015), 1501.02717.

\bibitem{2016ApJ...828...77V}
J.~H. {VanLandingham}, M.~C. {Miller}, D.~P. {Hamilton}, and D.~C.
  {Richardson},
\newblock \apj {\bf 828}, 77 (2016), 1604.04948.

\bibitem{2016ApJ...831..187A}
F.~{Antonini} and F.~A. {Rasio},
\newblock \apj {\bf 831}, 187 (2016), 1606.04889.

\bibitem{2016MNRAS.460.3494S}
A.~P. {Stephan} {\em et~al.},
\newblock \mnras {\bf 460}, 3494 (2016), 1603.02709.

\bibitem{2017arXiv170609896H}
B.-M. {Hoang}, S.~{Naoz}, B.~{Kocsis}, F.~A. {Rasio}, and F.~{Dosopoulou},
\newblock ArXiv e-prints  (2017), 1706.09896.

\bibitem{2018ApJ...865....2H}
A.~S. {Hamers}, B.~{Bar-Or}, C.~{Petrovich}, and F.~{Antonini},
\newblock \apj {\bf 865}, 2 (2018), 1805.10313.

\bibitem{Loeb:2016}
A.~{Loeb},
\newblock \apjl {\bf 819}, L21 (2016), 1602.04735.

\bibitem{Woosley:2016}
S.~E. {Woosley},
\newblock \apjl {\bf 824}, L10 (2016), 1603.00511.

\bibitem{Janiuk+2017}
A.~{Janiuk}, M.~{Bejger}, S.~{Charzy{\'n}ski}, and P.~{Sukova},
\newblock ArXiv e-prints {\bf 51}, 7 (2017), 1604.07132.

\bibitem{DOrazioLoeb:2017}
D.~J. {D'Orazio} and A.~{Loeb},
\newblock \prd {\bf 97}, 083008 (2018), 1706.04211.

\bibitem{2016PhRvL.116t1301B}
S.~{Bird} {\em et~al.},
\newblock Physical Review Letters {\bf 116}, 201301 (2016), 1603.00464.

\bibitem{2016PhRvD..94h4013C}
I.~{Cholis} {\em et~al.},
\newblock \prd {\bf 94}, 084013 (2016), 1606.07437.

\bibitem{2016PhRvL.117f1101S}
M.~{Sasaki}, T.~{Suyama}, T.~{Tanaka}, and S.~{Yokoyama},
\newblock Physical Review Letters {\bf 117}, 061101 (2016), 1603.08338.

\bibitem{2016PhRvD..94h3504C}
B.~{Carr}, F.~{K{\"u}hnel}, and M.~{Sandstad},
\newblock \prd {\bf 94}, 083504 (2016), 1607.06077.

\bibitem{2016ApJ...832L...2R}
C.~L. {Rodriguez}, M.~{Zevin}, C.~{Pankow}, V.~{Kalogera}, and F.~A. {Rasio},
\newblock \apjl {\bf 832}, L2 (2016), 1609.05916.

\bibitem{2018PhRvD..98h3007N}
K.~K.~Y. {Ng} {\em et~al.},
\newblock \prd {\bf 98}, 083007 (2018), 1805.03046.

\bibitem{2017PhRvD..95b4038H}
E.~A. {Huerta} {\em et~al.},
\newblock \prd {\bf 95}, 024038 (2017), 1609.05933.

\bibitem{2018ApJ...855...34G}
L.~{Gond{\'a}n}, B.~{Kocsis}, P.~{Raffai}, and Z.~{Frei},
\newblock \apj {\bf 855}, 34 (2018), 1705.10781.

\bibitem{2018PhRvD..97b4031H}
E.~A. {Huerta} {\em et~al.},
\newblock \prd {\bf 97}, 024031 (2018), 1711.06276.

\bibitem{2017ApJ...834..200M}
Y.~{Meiron}, B.~{Kocsis}, and A.~{Loeb},
\newblock \apj {\bf 834}, 200 (2017), 1604.02148.

\bibitem{2018arXiv180505335R}
L.~{Randall} and Z.-Z. {Xianyu},
\newblock arXiv e-prints , arXiv:1805.05335 (2018), 1805.05335.

\bibitem{2017arXiv170601385F}
W.~M. {Farr} {\em et~al.},
\newblock ArXiv e-prints  (2017), 1706.01385.

\bibitem{2018MNRAS.tmp.2223S}
J.~{Samsing} and D.~J. {D'Orazio},
\newblock \mnras  (2018), 1804.06519.

\bibitem{2018MNRAS.481.4775D}
D.~J. {D'Orazio} and J.~{Samsing},
\newblock \mnras {\bf 481}, 4775 (2018), 1805.06194.

\bibitem{2019PhRvD..99f3003K}
K.~{Kremer} {\em et~al.},
\newblock \prd {\bf 99}, 063003 (2019), 1811.11812.

\bibitem{2019PhRvD..99f3006S}
J.~{Samsing} and D.~J. {D'Orazio},
\newblock \prd {\bf 99}, 063006 (2019), 1807.08864.

\bibitem{2019arXiv190607189S}
J.~{Samsing}, A.~S. {Hamers}, and J.~G. {Tyles},
\newblock arXiv e-prints , arXiv:1906.07189 (2019), 1906.07189.

\bibitem{2006ApJ...640..156G}
K.~G{\"u}ltekin, M.~C. Miller, and D.~P. Hamilton,
\newblock \apj {\bf 640}, 156 (2006).

\bibitem{2014ApJ...784...71S}
J.~{Samsing}, M.~{MacLeod}, and E.~{Ramirez-Ruiz},
\newblock \apj {\bf 784}, 71 (2014), 1308.2964.

\bibitem{2017ApJ...840L..14S}
J.~{Samsing} and E.~{Ramirez-Ruiz},
\newblock \apjl {\bf 840}, L14 (2017), 1703.09703.

\bibitem{2018MNRAS.476.1548S}
J.~{Samsing} and T.~{Ilan},
\newblock \mnras {\bf 476}, 1548 (2018), 1706.04672.

\bibitem{2018ApJ...853..140S}
J.~{Samsing}, M.~{MacLeod}, and E.~{Ramirez-Ruiz},
\newblock \apj {\bf 853}, 140 (2018), 1706.03776.

\bibitem{2019MNRAS.482...30S}
J.~{Samsing} and T.~{Ilan},
\newblock \mnras {\bf 482}, 30 (2019), 1709.01660.

\bibitem{2018PhRvD..97j3014S}
J.~{Samsing},
\newblock \prd {\bf 97}, 103014 (2018), 1711.07452.

\bibitem{2018ApJ...855..124S}
J.~{Samsing}, A.~{Askar}, and M.~{Giersz},
\newblock \apj {\bf 855}, 124 (2018), 1712.06186.

\bibitem{2019ApJ...871...91Z}
M.~{Zevin}, J.~{Samsing}, C.~{Rodriguez}, C.-J. {Haster}, and
  E.~{Ramirez-Ruiz},
\newblock \apj {\bf 871}, 91 (2019), 1810.00901.

\bibitem{2018PhRvD..98l3005R}
C.~L. {Rodriguez} {\em et~al.},
\newblock \prd {\bf 98}, 123005 (2018), 1811.04926.

\bibitem{2011ApJ...741...82T}
T.~A. Thompson,
\newblock \apj {\bf 741}, 82 (2011).

\bibitem{2018ApJ...864..134R}
L.~{Randall} and Z.-Z. {Xianyu},
\newblock \apj {\bf 864}, 134 (2018), 1802.05718.

\bibitem{2018arXiv181110627F}
G.~{Fragione}, E.~{Grishin}, N.~W.~C. {Leigh}, H.~B. {Perets}, and R.~{Perna},
\newblock arXiv e-prints , arXiv:1811.10627 (2018), 1811.10627.

\bibitem{2019arXiv190208604R}
L.~{Randall} and Z.-Z. {Xianyu},
\newblock arXiv e-prints , arXiv:1902.08604 (2019), 1902.08604.

\bibitem{2019arXiv190309160F}
G.~{Fragione}, N.~{Leigh}, and R.~{Perna},
\newblock arXiv e-prints , arXiv:1903.09160 (2019), 1903.09160.

\bibitem{2019MNRAS.486.4443F}
G.~{Fragione} and A.~{Loeb},
\newblock \mnras {\bf 486}, 4443 (2019), 1903.10511.

\bibitem{2019arXiv190309659F}
G.~{Fragione} and O.~{Bromberg},
\newblock arXiv e-prints , arXiv:1903.09659 (2019), 1903.09659.

\bibitem{2019MNRAS.486.4781F}
G.~{Fragione} and B.~{Kocsis},
\newblock \mnras {\bf 486}, 4781 (2019), 1903.03112.

\bibitem{Kocsis:2012ja}
B.~Kocsis and J.~Levin,
\newblock Phys. Rev. D {\bf 85}, 123005 (2012).

\bibitem{2018ApJ...860....5G}
L.~{Gond{\'a}n}, B.~{Kocsis}, P.~{Raffai}, and Z.~{Frei},
\newblock \apj {\bf 860}, 5 (2018), 1711.09989.

\bibitem{2019ApJ...877...56L}
J.~{Lopez}, Martin, A.~{Batta}, E.~{Ramirez-Ruiz}, I.~{Martinez}, and
  J.~{Samsing},
\newblock \apj {\bf 877}, 56 (2019), 1812.01118.

\bibitem{2019arXiv190102889S}
J.~{Samsing} {\em et~al.},
\newblock arXiv e-prints , arXiv:1901.02889 (2019), 1901.02889.

\bibitem{2019arXiv190406353K}
K.~{Kremer}, W.~{Lu}, C.~L. {Rodriguez}, M.~{Lachat}, and F.~{Rasio},
\newblock arXiv e-prints , arXiv:1904.06353 (2019), 1904.06353.

\bibitem{2014LRR....17....2B}
L.~{Blanchet},
\newblock Living Reviews in Relativity {\bf 17} (2014), 1310.1528.

\bibitem{2017ApJ...846...36S}
J.~{Samsing}, M.~{MacLeod}, and E.~{Ramirez-Ruiz},
\newblock \apj {\bf 846}, 36 (2017), 1609.09114.

\bibitem{2018MNRAS.481.5436S}
J.~{Samsing}, N.~W.~C. {Leigh}, and A.~A. {Trani},
\newblock \mnras {\bf 481}, 5436 (2018), 1803.08215.

\bibitem{2019arXiv190409624H}
A.~S. {Hamers} and J.~{Samsing},
\newblock arXiv e-prints , arXiv:1904.09624 (2019), 1904.09624.

\bibitem{2019arXiv190608666H}
A.~S. {Hamers} and J.~{Samsing},
\newblock arXiv e-prints , arXiv:1906.08666 (2019), 1906.08666.

\bibitem{2019arXiv190201344H}
C.~{Hamilton} and R.~R. {Rafikov},
\newblock arXiv e-prints , arXiv:1902.01344 (2019), 1902.01344.

\bibitem{2019arXiv190201345H}
C.~{Hamilton} and R.~R. {Rafikov},
\newblock arXiv e-prints , arXiv:1902.01345 (2019), 1902.01345.

\bibitem{2019arXiv190700994H}
C.~{Hamilton} and R.~R. {Rafikov},
\newblock arXiv e-prints , arXiv:1907.00994 (2019), 1907.00994.

\bibitem{2011CQGra..28i4013H}
S.~{Hild} {\em et~al.},
\newblock Classical and Quantum Gravity {\bf 28}, 094013 (2011), 1012.0908.

\bibitem{2017CQGra..34d4001A}
B.~P. {Abbott} {\em et~al.},
\newblock Classical and Quantum Gravity {\bf 34}, 044001 (2017), 1607.08697.

\bibitem{Hansen:1972il}
R.~Hansen,
\newblock Phys. Rev. D {\bf 5}, 1021 (1972).

\bibitem{Lee:1993dt}
M.~H. Lee,
\newblock \apj {\bf 418}, 147 (1993).

\bibitem{2019ApJ...881...20R}
A.~{Rasskazov} and B.~{Kocsis},
\newblock \apj {\bf 881}, 20 (2019), 1902.03242.

\bibitem{2017ApJ...842L...2C}
X.~{Chen} and P.~{Amaro-Seoane},
\newblock \apjl {\bf 842}, L2 (2017), 1702.08479.

\bibitem{2018arXiv181106473A}
A.~{Askar}, A.~{Askar}, M.~{Pasquato}, and M.~{Giersz},
\newblock ArXiv e-prints  (2018), 1811.06473.

\bibitem{2018MNRAS.478.1844A}
A.~{Askar}, M.~{Arca Sedda}, and M.~{Giersz},
\newblock \mnras {\bf 478}, 1844 (2018), 1802.05284.

\bibitem{2011CQGra..28i4011K}
S.~{Kawamura} {\em et~al.},
\newblock Classical and Quantum Gravity {\bf 28}, 094011 (2011).

\bibitem{2018arXiv180206977I}
S.~{Isoyama}, H.~{Nakano}, and T.~{Nakamura},
\newblock ArXiv e-prints  (2018), 1802.06977.

\bibitem{TianQin}
J.~{Luo} {\em et~al.},
\newblock Classical and Quantum Gravity {\bf 33}, 035010 (2016), 1512.02076.

\bibitem{2015ApJ...800....9M}
M.~{Morscher}, B.~{Pattabiraman}, C.~{Rodriguez}, F.~A. {Rasio}, and
  S.~{Umbreit},
\newblock \apj {\bf 800}, 9 (2015), 1409.0866.

\bibitem{Heggie:1975uy}
D.~C. Heggie,
\newblock \mnras {\bf 173}, 729 (1975).

\bibitem{2015ApJ...808L..25G}
A.~M. {Geller} and N.~W.~C. {Leigh},
\newblock \apjl {\bf 808}, L25 (2015), 1506.08830.

\bibitem{2006ApJ...648..411K}
B.~{Kocsis}, M.~E. {G{\'a}sp{\'a}r}, and S.~{M{\'a}rka},
\newblock \apj {\bf 648}, 411 (2006), astro-ph/0603441.

\bibitem{2018ApJ...866L...5R}
C.~L. {Rodriguez} and A.~{Loeb},
\newblock \apjl {\bf 866}, L5 (2018), 1809.01152.

\bibitem{Spitzer:1969jx}
L.~J. Spitzer,
\newblock \apj {\bf 158}, L139 (1969).

\bibitem{1911MNRAS..71..460P}
H.~C. {Plummer},
\newblock \mnras {\bf 71}, 460 (1911).

\bibitem{2015PhRvD..92d4034S}
B.~{Sun}, Z.~{Cao}, Y.~{Wang}, and H.-C. {Yeh},
\newblock \prd {\bf 92}, 044034 (2015).

\bibitem{Wen:2003bu}
L.~Wen,
\newblock \apj {\bf 598}, 419 (2003).

\bibitem{2017CQGra..34m5011L}
N.~{Loutrel} and N.~{Yunes},
\newblock Classical and Quantum Gravity {\bf 34}, 135011 (2017), 1702.01818.

\bibitem{2016PhRvL.116w1102S}
A.~{Sesana},
\newblock Physical Review Letters {\bf 116}, 231102 (2016), 1602.06951.

\end{thebibliography}

\end{document}